\documentclass[%
  twocolumn,%
  unsortedaddress,%
  showpacs,%
  lengthcheck,%
  letterpaper,%
  amsfonts,%
  prb,%
  floatfix%
]{revtex4}

\usepackage[nohyperlinks]{acronym}
\usepackage{subfigure,graphicx}
\usepackage{amsmath,amssymb,amsxtra,bm,bbm}
\usepackage{fix-cm,type1cm,color} 
\usepackage{hyperref}
  \hypersetup{%
    breaklinks = {true},
    citecolor = {blue},
    colorlinks = {true},
    linkcolor = {red},
    pdfauthor = {\textcopyright\ Vitor M. Pereira},
    pdfcreator = {\LaTeX\ and \flqq hyperref\frqq},
    pdffitwindow = {true},
    pdfmenubar = {true},
    pdfpagelayout = {SinglePage},
    pdfstartview = {Fit},
    pdftoolbar = {true},
    plainpages = {false},
  }

\newcommand{\abs}[1]{\left| #1 \right|}
\newcommand{\ket}[1]{\vert #1 \rangle}
\newcommand{\average}[1]{\left\langle #1 \right\rangle}
\newcommand{\Ri}{\ensuremath{\vec{R}_i\,}}
\newcommand{\coloronline}{(Color online) }
\newcommand{\identity}{\mathbbm{1}}
\newcommand{\bG}{\bm{G}}
\newcommand{\bT}{\bm{T}}
\newcommand{\bV}{\bm{V}}

\DeclareMathOperator{\rank}{rank}
\DeclareMathOperator{\nullity}{nullity}

\begin{document}


\title{Modeling disorder in graphene}



\author{Vitor M. Pereira}
\affiliation{Department of Physics, Boston University, 590 
Commonwealth Avenue, Boston, MA 02215, USA}
\altaffiliation{The work reported here has been undertaken during a time 
at which V.M.P was simultaneously affiliated with 
CFP and Departamento de F{\'\i}sica, Faculdade de Ci\^encias
Universidade de Porto, 4169-007 Porto, Portugal.}

\author{J. M. B. Lopes~dos~Santos}
\affiliation{CFP and Departamento de F{\'\i}sica, Faculdade de Ci\^encias
Universidade de Porto, 4169-007 Porto, Portugal}

\author{A.~H. Castro Neto}
\affiliation{Department of Physics, Boston University, 590 
Commonwealth Avenue, Boston, MA 02215, USA}

\date{\today}


\begin{abstract}
We present a study of different models of local disorder in graphene. 
Our focus is on  the main effects that vacancies --- random, compensated and
uncompensated ---, local impurities and substitutional impurities bring into
the electronic structure of graphene. By exploring these types of
disorder and their connections,  we show that they introduce dramatic changes
in the low energy spectrum of graphene, viz. localized zero modes, strong
resonances, gap and pseudogap behavior, and non-dispersive midgap zero modes.
\end{abstract}

\maketitle



Graphene is poised to become a new paradigm in solid state physics and
materials science, owing to its truly bi-dimensional character and a host of
rich and unexpected phenomena\cite{Geim:2007,CastroNeto:2006b,CastroNeto:2007}. 
These have cascaded into the literature in
the wake of the seminal experiments that presented a relatively easy route
towards the isolation of graphene crystals\cite{Novoselov:2004}.

Carbon is a very interesting element, on account of
its chemical versatility: it can form more compounds than any other element
\cite{Chang:1991}. Its valence orbitals are known to hybridize in many
different forms like $sp^1$, $sp^2$, $sp^3$, and others. As a consequence,
carbon can exist in many stable allotropic forms, characterized by the different
relative orientions of the carbon atoms. Carbon binds through covalence, and
leads to the strongest chemical bonds found in nature. 
Common to the most interesting forms of carbon is the so-called graphene sheet,
a single plane of $sp^2$ carbon organized in an honeycomb lattice
(Fig.~\ref{fig:Honeycomb}). Graphite, for instance, is made of
stackings of graphene planes, nanotubes from rolled graphene sheets, and
fullerenes are wrapped graphene. Yet, for many years, it was believed that
graphene itself would be thermodynamically unstable. This presumption has been
overturned by a series of remarkable experiments in which truly bi-dimensional
(one atom thick) sheets of graphene have been isolated and characterized
\cite{Novoselov:2004}. This means that 
studies of the 2D (Dirac) electron gas can now be performed on a truly 2D 
crystal, as opposed to the traditional measurements made at interfaces as 
in \acs{MOSFET} and other structures \cite{Novoselov:2005b}. 

The crystalline simplicity of graphene --- a plane of $sp^2$ hybridized carbon
atoms arranged in a honeycomb lattice --- is deceiving. The
characteristics of the honeycomb lattice make graphene a half-filled
system with a \ac{DOS} that vanishes linearly at the neutrality point, and an
effective, low energy quasiparticle spectrum characterized by a dispersion
which is linear in momentum\cite{Wallace:1947} close to the Fermi energy.
These two features underlie the
unconventional electronic properties of this material, whose quasiparticles
behave as Dirac massless chiral electrons\cite{Semenoff:1984}. Consequently,
 many phenomena of the realm of \ac{QED} find a
practical realization in this solid state material. They
include:  the minimum conductivity when the carrier density tends to
zero\cite{Novoselov:2005}; 
the new half-integer quantum Hall effect, measurable up to room
temperature\cite{Novoselov:2005}; 
Klein tunneling\cite{Katsnelson:2007}; 
strong overcritical positron-like resonances in the Coulomb scattering
cross-section analogous to supercritical nuclei in \ac{QED}
\cite{Pereira:2007,Shytov:2007};
the zitterbewegung in confined structures\cite{Peres:2006b}; 
anomalous Andreev reflections\cite{Miao:2007,Beenakker:2007}; 
negative refraction \cite{Cheianov:2007} in p-n junctions.

Arguably, the most interesting and promising properties from the technological
point of view are its great crystalline quality, high
mobility and resilience to very high current densities\cite{Geim:2007}; the
ability to tune the carrier density through a
gate voltage\cite{Novoselov:2004}; the absence of
backscattering\cite{Ando:1998b} and the fact that graphene exhibits both spin
and valley degrees of freedom which might be harnessed in envisaged
spintronic\cite{Kane:2007b,Cho:2007} or valleytronic devices\cite{Rycerz:2007}.

Disorder, ever present in graphene owing to its exposed surface and the
substrates, is the central concern of this paper. In particular, we focus on the
effects of vacancies and random impurities in the electronic structure of bulk
graphene. The models examined below apply to situations in which Carbon atoms
are extracted from the graphene plane (e.g. through
irradiation\cite{Esquinazi:2003}), in which
adatoms and/or adsorbed species attach to the graphene plane\cite{Schedin:2007},
or in which some carbon atoms are chemically substituted for other
elements. They are, therefore, models of local disorder. 
We do not consider
explicitly other sources of disorder like rough edges or
ripples\cite{Meyer:2007}, or the dramatic effects of Coulomb impurities, which  
have been discussed elsewhere\cite{Pereira:2007,Lewenkopf:2007}.
In this article we expand the discussion of
vacancies initiated in Ref.~\onlinecite{Pereira:2005}, using the same
techniques, and discuss the consequences of local disorder originally presented
in Ref.~\onlinecite{Pereira:2006PhD}. Numerically we resort to exact
diagonalization calculations and to the recursion
method\cite{Haydock:1972a,Haydock:1975a}. The latter allows the calculation of
the \ac{DOS} and other spectral quantities for very large system sizes with
disorder. In our case, the calculations below refer to honeycomb lattices
with $4\times10^6$ carbon atoms, a size already of the order of magnitude of
the real samples, if not larger for some experiments.

The article is organized as follows. In section \ref{sec:Honeycomb} we
present the basic electronic properties of electrons in the honeycomb 
lattice, mostly to introduce the notation and the details relevant for the
subsequent discussions. In section \ref{sec:Disorder} we present our results
regarding the different models of disorder. This section is subdivided according
to the different models of disorder studied: vacancies in
\ref{subsec:Disorder-Vacancies} and \ref{subsec:SelectiveDilution}, local
impurities in \ref{subsec:Disorder-LocalImpurities} and substitutional
impurities in \ref{subsec:Disorder-SubstImpurities}. The discussion of the
results is kept within each subsection and the principal findings of this paper
are highlighted in the conclusion, in section \ref{sec:Conclusions}.


\section{Electrons in a Honeycomb Lattice \label{sec:Honeycomb}}

\begin{figure}
  \begin{center}
    \subfigure[][]{%
      \includegraphics*[width=0.5\columnwidth]{%
        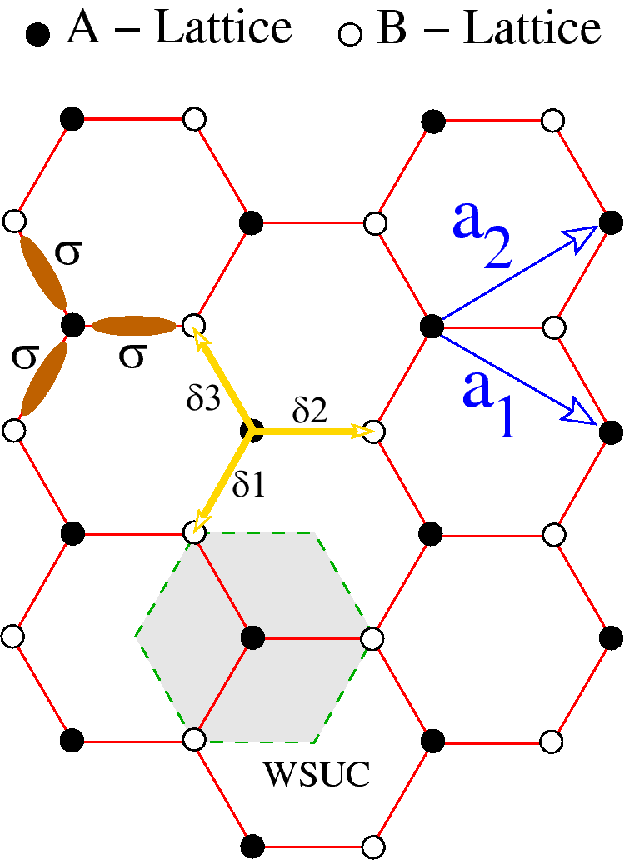}%
      \label{fig:Honeycomb}%
    }%
    \quad%
    \subfigure[][]{%
      \includegraphics*[bb=0 -20 107 129,width=0.4\columnwidth]{%
        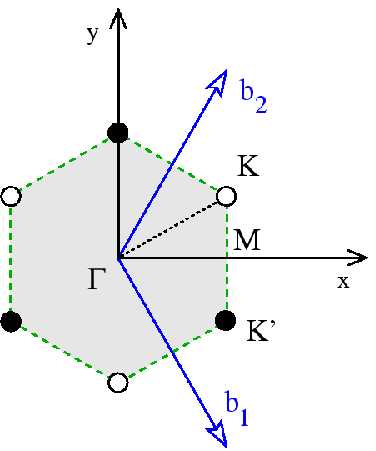}%
      \label{fig:HoneycombBZ}%
    }%
  \end{center}
  \caption{
    \coloronline
    In \subref{fig:Honeycomb} $\vec{a}_1$ and $\vec{a}_2$ are the
    primitive vectors that define 
    the \ac{WS} unit cell highlighted as the dashed hexagon. 
    The lattice parameter, $a$, is $\simeq 1.4$ \AA.
    The first \ac{BZ} of the associated reciprocal lattice is shown in
    (b), together with the points of high symmetry
    $\Gamma$, $M$ and the two nonequivalent $K$ and $K'$. 
  }
\end{figure}
%

Graphene consists of carbon atoms organized into a honeycomb lattice, bonded
through covalence between two $sp^2$ orbitals of neighboring atoms
(Fig.~\ref{fig:Honeycomb}). The graphene plane is defined by the plane
of the $sp^2$ orbitals. The saturation of the resulting $\sigma$ bonding
orbitals, leaves an extra electron at the remaining $2p_z$ orbital per carbon
atom. Ideal graphene has therefore a half-filled electronic ground state.

The Bravais lattice that underlies the translation symmetries of the honeycomb
lattice is the triangular lattice, whose primitive vectors $\vec{a}_1$ and
$\vec{a}_2$ are depicted in Fig.~\ref{fig:Honeycomb}. One of the
consequences is the existence of two atoms per unit cell, that define two
sublattices ($A$ and $B$ in the figure): indeed, the honeycomb lattice can be
thought as two interpenetrating triangular lattices. This bipartite nature of
the crystal lattice, added to the half-filled band, imposes an important
particle-hole symmetry as will be discussed later.
 
The electronic structure of graphene can be captured within a tight-binding
approach, in which the electrons are allowed to hop between immediate neighbors
with hopping integral $t\simeq 2.7$~eV, and also between next-nearest neighbors
with an additional hopping $t^\prime$:
\begin{equation}
  H = - t \sum_{\langle i,j \rangle} c^\dagger_i c_j
      - t' \sum_{\langle\langle i,j \rangle\rangle} c^\dagger_i c_j + \text{
h.c. }
  \label{eq:Hamiltonian}
  \,.  
\end{equation}
The presence of the second term introduces an asymmetry between the valence and
conduction bands, thus violating particle-hole symmetry. 
To emphasize the two sublattice structure of the honeycomb, we can write the 
Hamiltonian as
\begin{equation}
\begin{split}
  H = &- t \sum_{i \in A, \delta} a^\dagger_i b_{i+\delta} 
       - t \sum_{i \in B, \delta} b^\dagger_i a_{i+\delta}\\
      &- t'\sum_{i \in A, \Delta} a^\dagger_i a_{i+\Delta}
       - t'\sum_{i \in B, \Delta} b^\dagger_i b_{i+\Delta}\,,
\end{split}
\label{eq:Hamiltonian-2}
\end{equation}
with operators $a_{i}$ and $b_{i}$ pertaining to sublattices $A$ and $B$
respectively. 
The vectors $\vec{\delta}$ connect atom $i$ to its immediate
neighbors, whereas the $\vec{\Delta}$ connect atom $i$ to its six second
neighbors. Fourier transforming eq.~\eqref{eq:Hamiltonian-2} 
and introducing a spinor notation for the sublattice amplitudes leads to 
\begin{equation}
  H = \sum_k \Psi^\dagger_k 
        \begin{pmatrix}
          \epsilon_2(k) & \epsilon_1(k) \\
          \epsilon_1(k)^* & \epsilon_2(k) 
        \end{pmatrix}
      \Psi_k  
  \,,\text{  with }
  \Psi_k  = \begin{pmatrix}
              a_{k} \\
              b_{k} 
            \end{pmatrix}
  \,.
  \label{eq:Hamiltonian-Kspace}
\end{equation}
Since the spin degree of freedom does not play a role in our discussion other
than through a degeneracy factor, it will been omitted, for simplicity.
The functions
$\epsilon_1(k)$ and $\epsilon_2(k)$ read%
\begin{gather}
  \epsilon_1(k) = -t \sum_{\vec{\delta}} e^{-i \vec{\delta}.\vec{k}} 
  \,,\qquad 
  \epsilon_2(k) = -t' \sum_{\vec{\Delta}} e^{-i \vec{\Delta}.\vec{k}}
  \\ 
  \begin{split}
  \epsilon_2(k) = & - 2t' \cos(\sqrt{3}k_ya) \\
                  & - 4t' \cos\left(\frac{\sqrt{3}}{2}k_ya\right)
                      \cos\left(\frac{3}{2}k_xa\right)
  \\
  \abs{\epsilon_1(k)}^2 = & 3t^2 + 2t^2\cos(\sqrt{3}k_ya) \\
                          & + 4t^2\cos\left(\frac{\sqrt{3}}{2}k_ya\right)
                              \cos\left(\frac{3}{2}k_xa\right)
  \end{split}
  \label{eq:Dispersions}
  \,,
\end{gather}
where $\epsilon_2(k)$ alone is the dispersion relation of a triangular lattice,
and yield, after diagonalization of \eqref{eq:Hamiltonian-Kspace}, the
dispersion relations for graphene:
\begin{equation}
  E_\pm(k) = \epsilon_2(k) \pm \abs{t} \sqrt{3 - \frac{\epsilon_2(k)}{t'}}
  \,. 
  \label{eq:Dispersion}
\end{equation}
%
%
%
%
\begin{figure}
  \centering
  \subfigure[][]{
    \includegraphics*[width=0.56\columnwidth]{%
      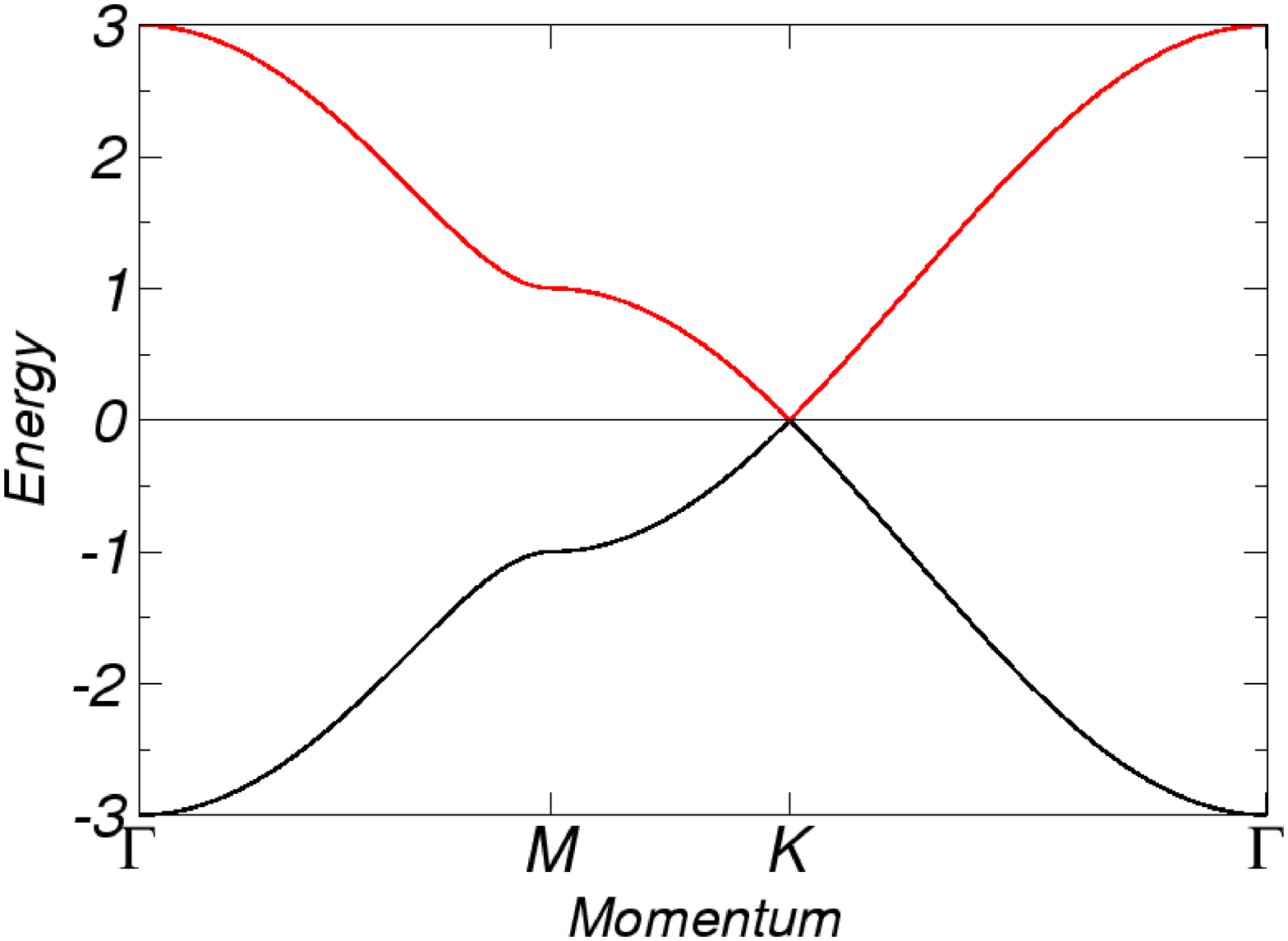}
    \label{fig:BandSelected}
  }%
  \subfigure[][]{%
    \includegraphics*[width=0.45\columnwidth]{%
      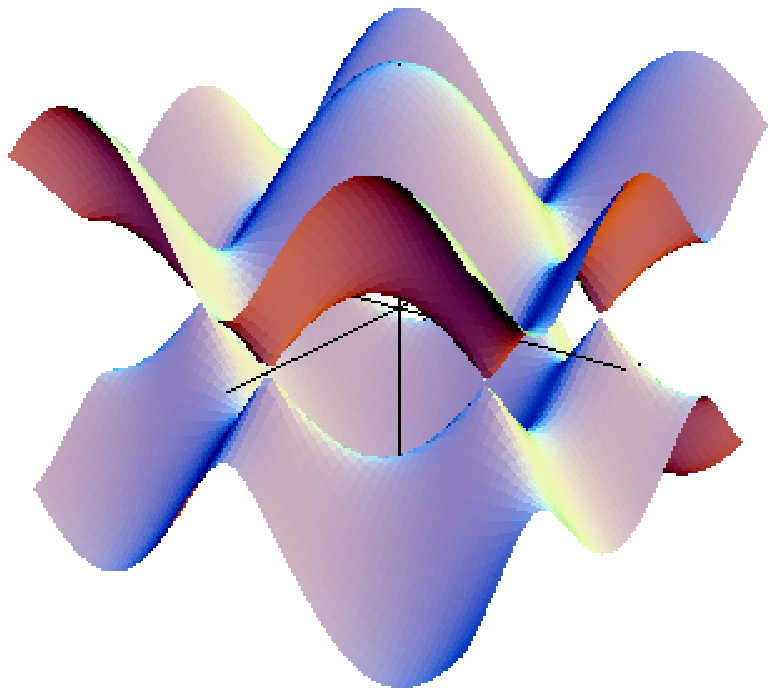}%
    \label{fig:HoneycombSpectrum}%
  }%
  \caption{
    \coloronline
    \subref{fig:BandSelected} Band structure along the symmetry
    directions of the reciprocal \ac{BZ} of the honeycomb lattice. 
    \subref{fig:HoneycombSpectrum} Band structure of
    graphene ($t'=0$) with the two bands touching at the $K$ and $K'$ 
    points of the \ac{BZ}.
  }
  \label{fig:Bands}
\end{figure}
%
The two bands $E_\pm(k)$ are represented in Fig.~\ref{fig:HoneycombSpectrum} in
the domain $k_{x,y} \in [-\pi,\pi]$. This unusual  bandstructure makes graphene
 very peculiar with valence and conduction
bands touching at the Fermi energy, at a set of points at the edge of the first
Brillouin zone, equivalent to the points $K$ and $K'$ by suitable reciprocal
lattice translations. Its low energy physics is dictated
by the dispersion
around those two inequivalent points, which turns out to be linear in $k$. 
In fact, expanding \eqref{eq:Dispersions} around either
\begin{equation}
  K  = \frac{4\pi}{3 \sqrt{3} a} \left(\frac{\sqrt{3}}{2}, \frac{1}{2} \right)
  \text{  or  }
  K' = \frac{4\pi}{3 \sqrt{3} a} \left(\frac{\sqrt{3}}{2}, -\frac{1}{2} \right)
\end{equation}
one gets the so-called $\vec{K}.\vec{p}$ \emph{effective}
bandstructure\cite{Wallace:1947}:
\begin{equation}
  E(\vec{K} + \vec{q}) = -3 t' \pm \nu_F \abs{\vec{q}} + \mathcal{O}(q^2)
  \label{eq:KpBands}
  \,,
\end{equation}
with a Fermi velocity, $v_F$ ($\hbar v_F \equiv \nu_F = 3ta/2$, and we take
units in which $\hbar=1$ and $t=1$). When $t'=0$ the 
dispersion is purely
conical, as in a relativistic electron in 2D. For this reason, the two cones
tipped at $K$ and $K'$ are known as Dirac cones. The low-energy,
continuum limit of \eqref{eq:Hamiltonian-2} is given by 
\begin{equation}
  H = v_F \int d^2\bm{r} \: \psi^\dagger(\bm{r}) \: \vec{\sigma}\cdot\vec{p} 
      \: \psi(\bm{r})
  \label{eq:Dirac}
  \,,
\end{equation}
where $\psi(\bm{r})$ is a two dimensional spinor obeying the Dirac equation in
2D \cite{Gonzalez:1992}.

\begin{figure}
  \centering
  \includegraphics*[width=\columnwidth]{%
    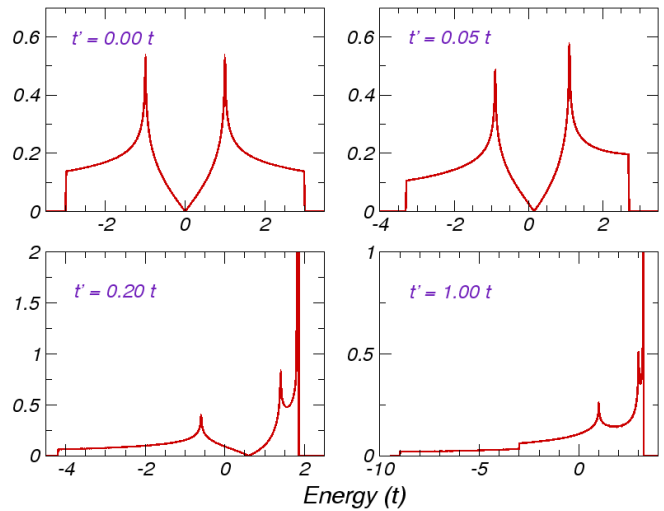}
  \caption{
    \coloronline
    \ac{DOS} associated with eq.~\eqref{eq:Dispersion} for 
    different values of the next nearest neighbor hopping $t'$. 
  }
  \label{fig:DOS-Comparison}
\end{figure}

Some quantitative aspects of graphene's band structure
\eqref{eq:Dispersion} are plotted in
Figs.~\ref{fig:Bands} and \ref{fig:DOS-Comparison}.
%
%
%
In panel \ref{fig:BandSelected} the band dispersion is plotted along
the symmetry directions of the \ac{BZ} indicated in
Fig.~\ref{fig:HoneycombBZ}, and in panel
\ref{fig:DOS-Comparison} the
\ac{DOS} for different values of the nearest-neighbor hopping, $t'$, are
plotted. Focusing on the particle-hole symmetric case ($t=0$), it is clear that,
besides the marked van~Hove singularities at $E=\pm t$, the most important
feature is the linear vanishing of the \ac{DOS} at the Fermi level, a fact that
is at the origin of many transport anomalies in this material
\cite{Peres:2005a,CastroNeto:2007}. 

Particle-hole symmetry in this problem arises from the bipartite nature of the
honeycomb lattice, and is a general property of systems whose underlying crystal
lattice has this nature. When we have a
bipartite lattice, the basis vectors of the Hilbert space can be ordered
so that, for any ket, $\ket{\varphi}$, the amplitudes in sublattice $A$ come
first. For example, if $\{\phi_A^1, \phi_A^2, \dots, \phi_A^N\}$ are the Wannier
functions for the orbitals in sublattice $A$, and $\{\phi_B^1, \phi_B^2, \dots,
\phi_B^N\}$ the ones in sublattice $B$, then our ordered basis could be
$\{\phi_A^1, \dots, \phi_A^N; \phi_B^1, \dots, \phi_B^N\}$. If the Hamiltonian
includes hopping only between nearest neighbors, this means that it only
promotes itinerancy between different sublattices. The stationary
Schr\"{o}dinger
equation then reads, in matrix block form in the ordered basis,
\begin{equation}
  \begin{pmatrix}
    0 & h_{AB} \\
    h_{AB}^\dagger & 0 \\
  \end{pmatrix}
  \begin{pmatrix}
    \varphi_A\\
    \varphi_B\\
  \end{pmatrix}
  = E
  \begin{pmatrix}
    \varphi_A\\
    \varphi_B\\
  \end{pmatrix}
  \label{eq:PholeSymm-1}
  \,. 
\end{equation}
Expanding we get 
\begin{equation}
  \begin{cases}
    & h_{AB} \; \varphi_B = E \varphi_A \\
    & h_{AB}^\dagger \; \varphi_A = E \varphi_B
  \end{cases}
  \Rightarrow
    (h_{AB}^\dagger h_{AB}) \; \varphi_B = E^2 \varphi_B \\
  \label{eq:PholeSymm-2}  
  \,,
\end{equation}
and therefore, if $E$ is an eigenstate, so is $-E$. For a half-filled system,
the elementary excitations around the Fermi sea can be thought, as usual, as
particle-hole pairs. Since in that case $E_F=0$, particles and holes have
symmetric dispersions. This is completely analogous to the situation found in
simple semiconductors or semimetals, although matters are slightly more
complicated in graphene because there are two degenerate points, $K$ and $K'$ in
the \ac{BZ}. Thus there will be two families of particle and hole excitations:
one associated with the Dirac cone at $K$, and the other with the cone at $K'$,
like in a multi-valley semiconductor.


\section{Local Disorder in Graphene\label{sec:Disorder}}

Disorder is present in any real material, graphene being no
exception. In fact,  true long-range order in 2D implies a broken
continuous symmetry (translation), which violates the Hohenberg-Mermin-Wagner
theorem \cite{Mermin:1966,Hohenberg:1967}. So, by this reason alone,  
defects must be present in
graphene and, in a sense, as paradoxical as it might sound, are presumably at
the basis of its thermodynamic stability. 

But the study of disorder effects on graphene is motivated by more extraordinary
experimental results. One of them is the study undertaken by
\citet{Esquinazi:2003} in which \ac{HOPG} samples were irradiated via high
energy proton beams. As a result, the experiments revealed that the samples
acquired a magnetic moment, displaying long range ferromagnetic order up to
temperatures much above $300$~K. This triggered enormous interest, since the
technological possibilities arising from organic magnets are many and varied.
Furthermore, carbon, being the most covalent of the elements, has a strong
tendency to saturate its shell in its allotropes, and is somehow the antithesis
of magnetism. Besides the moment formation, it was found that the magnitude of
the saturation moment registered in hysteresis curves was progressively
increased with successive irradiations. This is strong evidence that the defects
induced by the proton beam are playing a major role in this magnetism. In this
context the study of defects and disorder in graphene gains a significant
pertinence.

In the following paragraphs we will unveil some details and peculiarities that
emerge from different models of disorder applied to free electrons in the
honeycomb lattice. 


\subsection{Vacancies \label{subsec:Disorder-Vacancies}}

Vacancies are one of the defects more likely to be induced in the graphene
structure by proton irradiation. 
A vacancy is simply the absence of an atom at a given site.
When an atom is removed two scenarios are possible: either the disrupted
bonds remain as dangling bonds, or the structure undergoes a bond reconstruction
in the vicinity of the vacancy, with several possible outcomes
\cite{Ding:2005}. In either case, a slight local distortion of the lattice is
expected. In the following discussion, however, it is  assumed that, as first
approximation, the creation of a vacancy has the sole effect of removing the
$\pi_z$ orbital at a lattice point, together with its conduction band electron.
In this sense, the physics of the conduction band electrons is still described
by the Hamiltonian \eqref{eq:Hamiltonian}, where now the hopping to the
vacancy sites is \emph{forbidden}. 

  
\subsubsection{Vacancies and a theorem
\label{subsubsec:Graphene-Disorder-Vacancies-Theorem}}

Vacancies have an interesting consequence when $t'=0$. If the distribution of
vacant sites is uneven between the two sublattices,
zero energy modes will necessarily appear. This follows from a theorem in linear
algebra\cite{Mudry:2002} and can be seen as follows. 
Assume, very generally, that we have a bipartite lattice, with sublattices $A$
and $B$ (It can be any bipartite lattice like the square or honeycomb lattices
in 2D, cubic in 3D, etc.), and that the number of orbitals/sites in $A$($B$) is
$N_A$($N_B$). 
Just as we did before, the basis vectors of the Hilbert space can 
always be ordered so that any ket, $\ket{\Psi}$, has the amplitudes on 
sublattice $A$ appearing first, followed by the amplitudes on sublattice $B$:
\begin{equation}
  \ket{\Psi} = ( \varphi_A, \varphi_B ) 
             = ( \phi_A^1, \phi_A^2, \dots, \phi_A^{N_A}; \phi_B^1, \phi_B^2,
\dots, \phi_B^{N_B})
  \,.
\end{equation}
We now consider an Hamiltonian containing only nearest-neighbor hopping, plus
some local energy ($\epsilon_A, \epsilon_B$) on each sublattice. The
corresponding stationary Schr\"{o}dinger equation will then be (in matrix block
form
that respects the ordering of the basis)
\begin{equation}
  \mathcal H \ket{\Psi} = E \ket{\Psi} \mapsto
  \begin{pmatrix}
    \epsilon_A \identity_{N_A} & h_{AB} \\
    h_{AB}^\dagger & \epsilon_B \identity_{N_B} \\
  \end{pmatrix}
  \begin{pmatrix}
    \varphi_A\\
    \varphi_B\\
  \end{pmatrix}
  = E
  \begin{pmatrix}
    \varphi_A\\
    \varphi_B\\
  \end{pmatrix}
  \label{eq:VacanyTheorem-1}
  \,, 
\end{equation}
where $\identity_M$ is the $M\times M$ identity matrix, $h_{AB}$ a $N_A \times
N_B$
matrix, and $\varphi_A$ ($\varphi_B$) a vector in a subspace of dimension $N_A$
($N_B$). 

To analyze the spectrum we note that 
\begin{equation}
  \begin{cases}
    h \: \varphi_B = (E-\varepsilon_A) \varphi_A \\
    h^\dagger \: \varphi_A = (E-\varepsilon_B) \varphi_B \\
  \end{cases}
  \label{eq:VacanyTheorem-2}
  \,,
\end{equation}
which, from cross-substitution, implies that
\begin{equation}
  h^\dagger h \varphi_B = (E-\varepsilon_A)(E-\varepsilon_B) \varphi_B
  \,.
\end{equation}
If we call $\lambda^2$ to the (non-negative) eigenvalues of $h^\dagger h$, the
spectrum of $\mathcal H$ is then
\begin{equation}
  E = \frac{\varepsilon_A + \varepsilon_B}{2} \pm 
      \sqrt{
        \Bigl(
          \frac{\varepsilon_A - \varepsilon_B}{2}
        \Bigr)
        + \lambda^2
      }
  \,.
\end{equation}
The symmetry about $(\varepsilon_A + \varepsilon_B)/2$ simply reflects the
particle-hole symmetry.


\subsubsection{Uncompensated lattices} 

States of a \emph{peculiar} nature should appear when the number of sites in
each sublattice is different. Without any loss of generality we take $N_A >
N_B$. Since the block $h_{AB}$ in \eqref{eq:VacanyTheorem-1} is a
linear application from a vector space having $\dim(A) = N_A$, onto a vector
space $B$ with $\dim(B) = N_B$, it follows from basic linear algebra that
\begin{itemize}
  \item $\rank(h_{AB}) = \rank(h_{AB}^\dagger) = N_B$;
  \item $h_{AB} \: \varphi_B = 0$ has no solutions other than the trivial one;
  \item $h_{AB}^\dagger \: \varphi_A = 0$ has non-trivial solutions that we call
$\varphi_A^0$. 
\end{itemize}
From the rank-nullity theorem,
\begin{equation}
  \rank(h_{AB}^\dagger) + \nullity(h_{AB}^\dagger) = N_A
  \,,  
\end{equation}
and hence the null space of $h_{AB}^\dagger$ has dimension:
$\nullity(h_{AB}^\dagger) = N_A-N_B$. Consequently, there are states of the form
\begin{equation*}
  \ket{\Psi^0} = (\varphi_A^0;0)
\end{equation*}
in which $\varphi_A^0$ satisfies $h_{AB}^\dagger \: \varphi_A^0 = 0$, that are
eigenstates of $\mathcal H$ with eigenvalue $\varepsilon_A$:
\begin{equation}
  \mathcal H \ket{\Psi^0} = E \ket{\Psi} \Leftrightarrow
  \begin{cases}
    h \: 0 = (\varepsilon_A-\varepsilon_A) \varphi_A \\
    h^\dagger \: \varphi_A^0 = (\varepsilon_A-\varepsilon_B) 0 \\
  \end{cases}
  \,.
\end{equation}
Furthermore, since $\nullity(h_{AB}^\dagger) = N_A-N_B$ implies the existence of
$N_A-N_B$ linearly independent $\varphi_A^0$, this eigenstate has a degeneracy
of $N_A-N_B$. It should be stressed that a state of the form $(\varphi_A;0)$ has
only amplitude in the $A$ sublattice. Therefore, we conclude that, whenever the
two sublattices are not balanced with respect to their number of atoms, there
will appear $N_A - N_B$ states with energy $E_A$, all linearly independent and
localized only on the \emph{majority} sublattice. In addition, one can modify
sublattice $B$ in any way (remove more sites, for instance) that
these zero modes will remain totally undisturbed.

We remark that in the above the details of the hopping matrix
$h_{AB}$ were not specified and need not be. The result holds in general,
provided that the hopping induces transitions between different sublattice only,
and that the diagonal energies are constant (diagonal disorder is excluded).


\subsubsection{Zero modes}

The case with $\varepsilon_A = \varepsilon_B = 0$ is of obvious relevance for
us, since our model for pristine graphene does not include any local potentials.
In this situation, the above results imply that introducing a vacancy in an
otherwise perfect lattice, immediately creates a zero energy mode. Now this is
important because those states are created precisely at the Fermi level, and
have this peculiar topological localization determining that they should live in
just one of the lattices. 

Even more interestingly, it is possible to obtain the exact analytical
wavefunction associated with the zero mode induced by a single vacancy in a
honeycomb lattice. This was done by the authors and collaborators in
Ref.~\onlinecite{Pereira:2005}, and will not be repeated here. We only
mention that the wavefunction can be constructed by an appropriate matching of
the zero modes of two semi-infinite and complementary ribbons of graphene, and
that,  in the continuum limit, the wavefunction of the zero mode introduced
by one vacancy has the form\cite{Pereira:2005}
\begin{eqnarray}
  \Psi(x,y) \simeq 
    \frac{e^{i \bm{K^\prime}.\bm{r}}}{x+iy} +
    \frac{e^{i \bm{K}.\bm{r}}}{x-iy}
  \,.
  \label{eq:psi_x_y}
\end{eqnarray}
The important point is that the amplitude of this state decays with the distance
to the vacancy as $\sim 1/r$, and thus has a quasi-localized character, although
strictly not-normalizable. And such quasi-localized state appears exactly at the
Fermi level.

Should another vacancy be introduced in the same sublattice, we already know
that another zero mode will appear. However, the nature of the two zero modes
will depend whether the vacancies are close or distant. In the latter case, the
hybridization between the two modes should be small on account of the $1/r$
decay, and we can expect two states of the form \eqref{eq:psi_x_y}
about each vacancy site. Of course significant effects in the thermodynamic
limit can only arise with a finite concentration of vacancies, and for such
analysis we undertook the numerical calculations described next.  
%
\begin{figure}
  \centering
  \subfigure[][]{%
     \includegraphics*[width=0.48\columnwidth]{%
      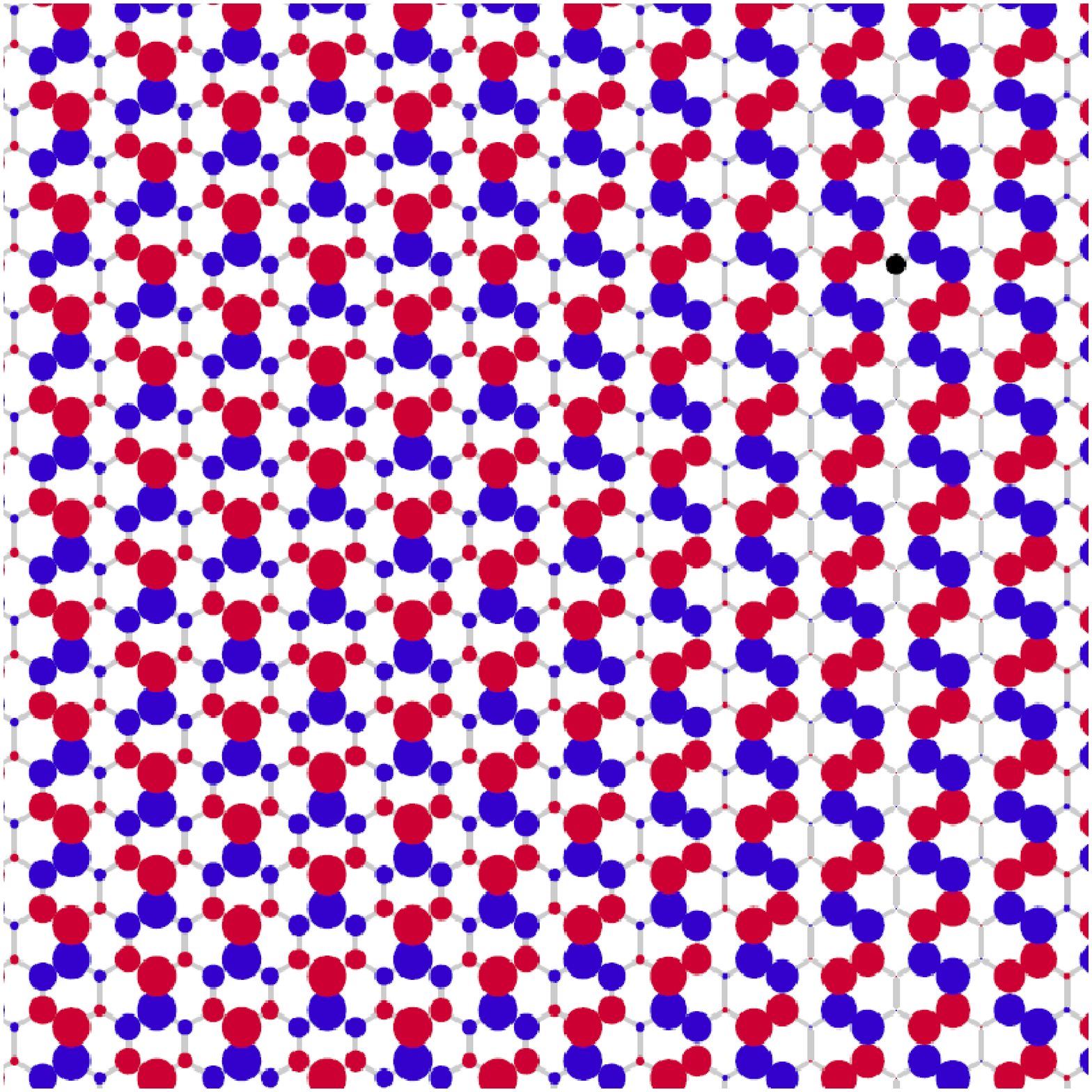}%
    \label{fig:Eigenstate-NearZero}%
  }%
  \quad%
  \subfigure[][]{%
    \includegraphics*[width=0.48\columnwidth]{%
      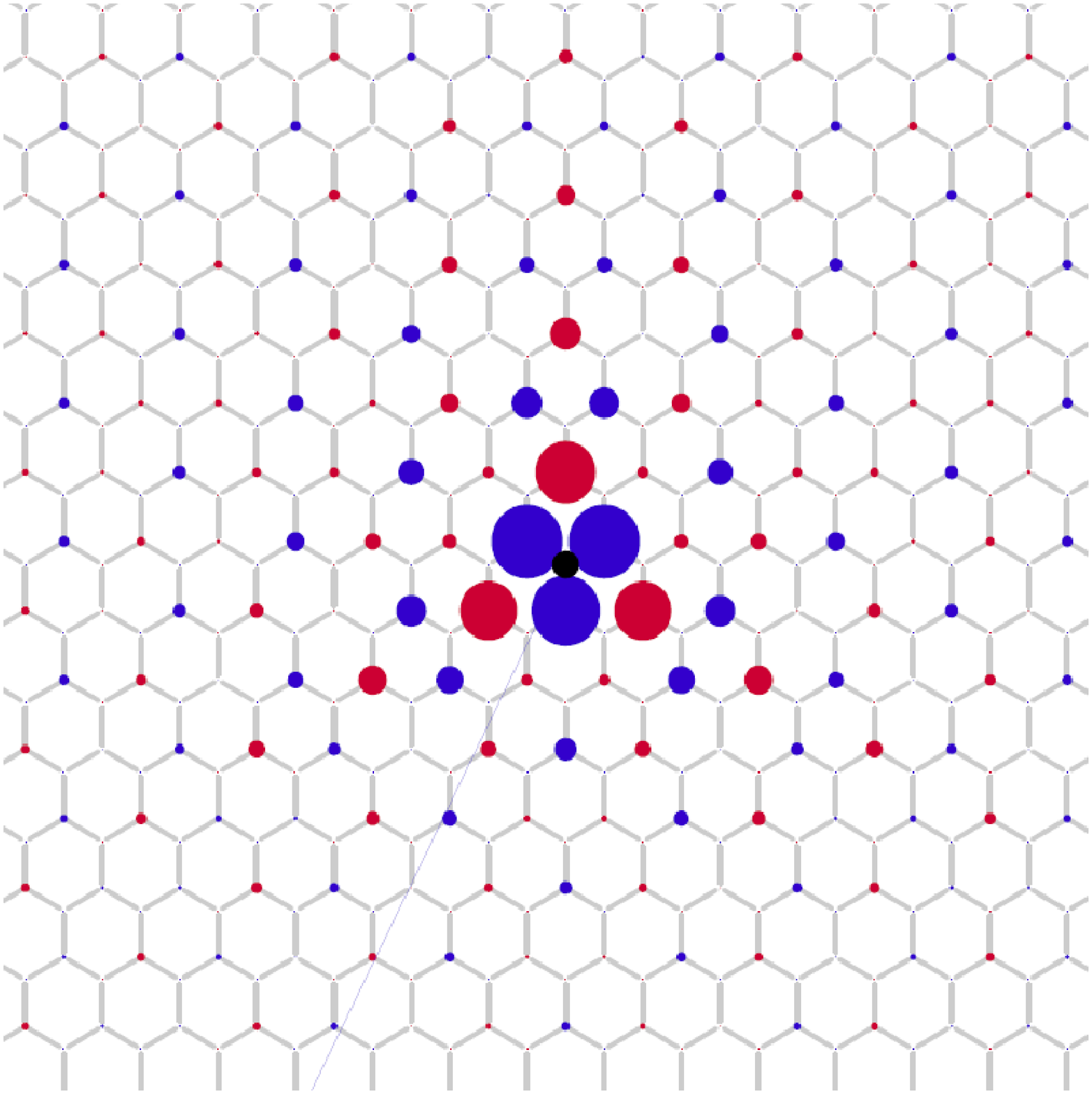}%
    \label{fig:Eigenstate-AtZero}%
  }%
  \\%
  \subfigure[][]{%
    \includegraphics*[width=0.48\columnwidth]{%
      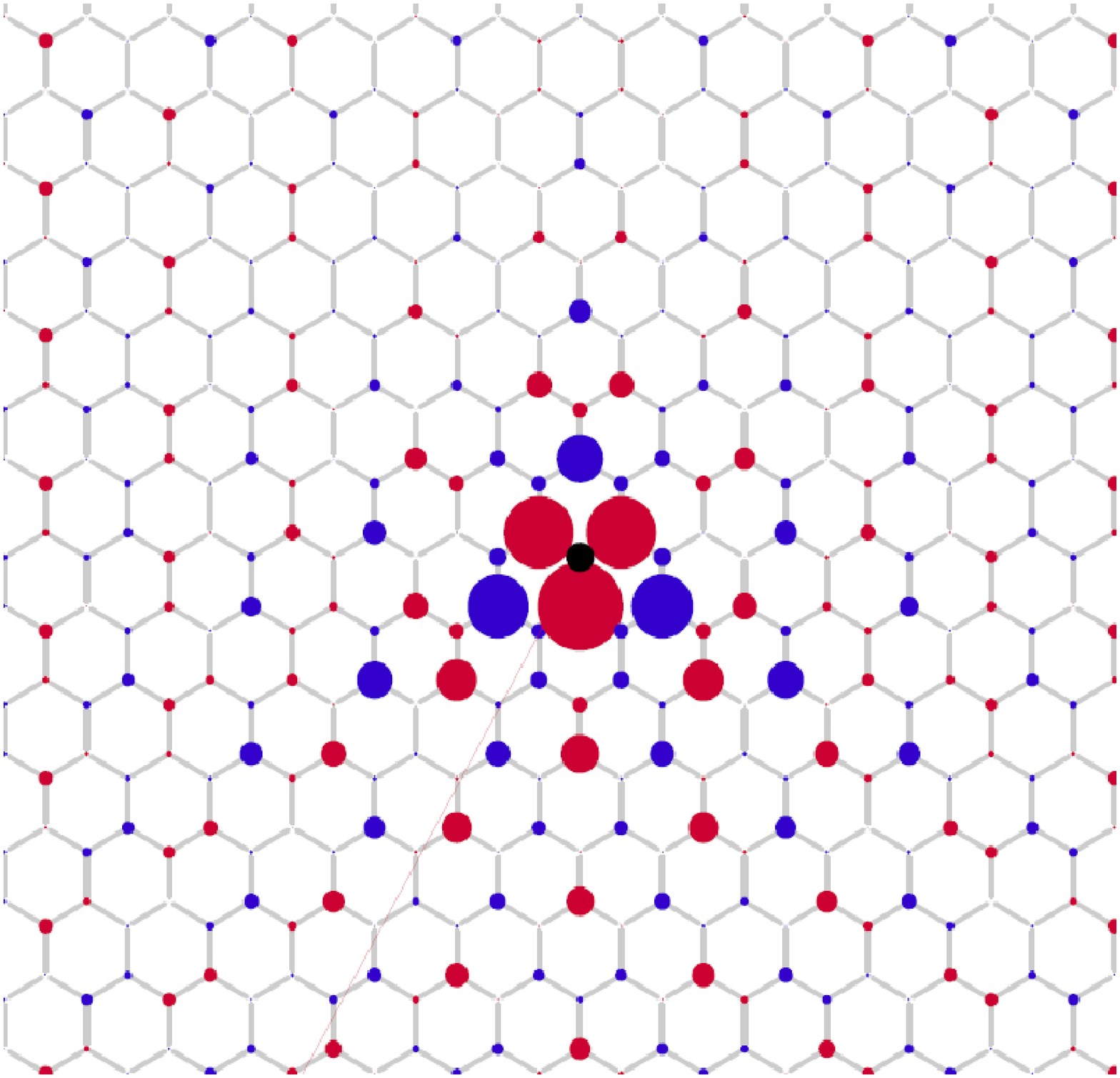}%
    \label{fig:Eigenstate-Tlinha-1}%
  }%
  \quad%
  \subfigure[][]{%
    \includegraphics*[width=0.48\columnwidth]{%
      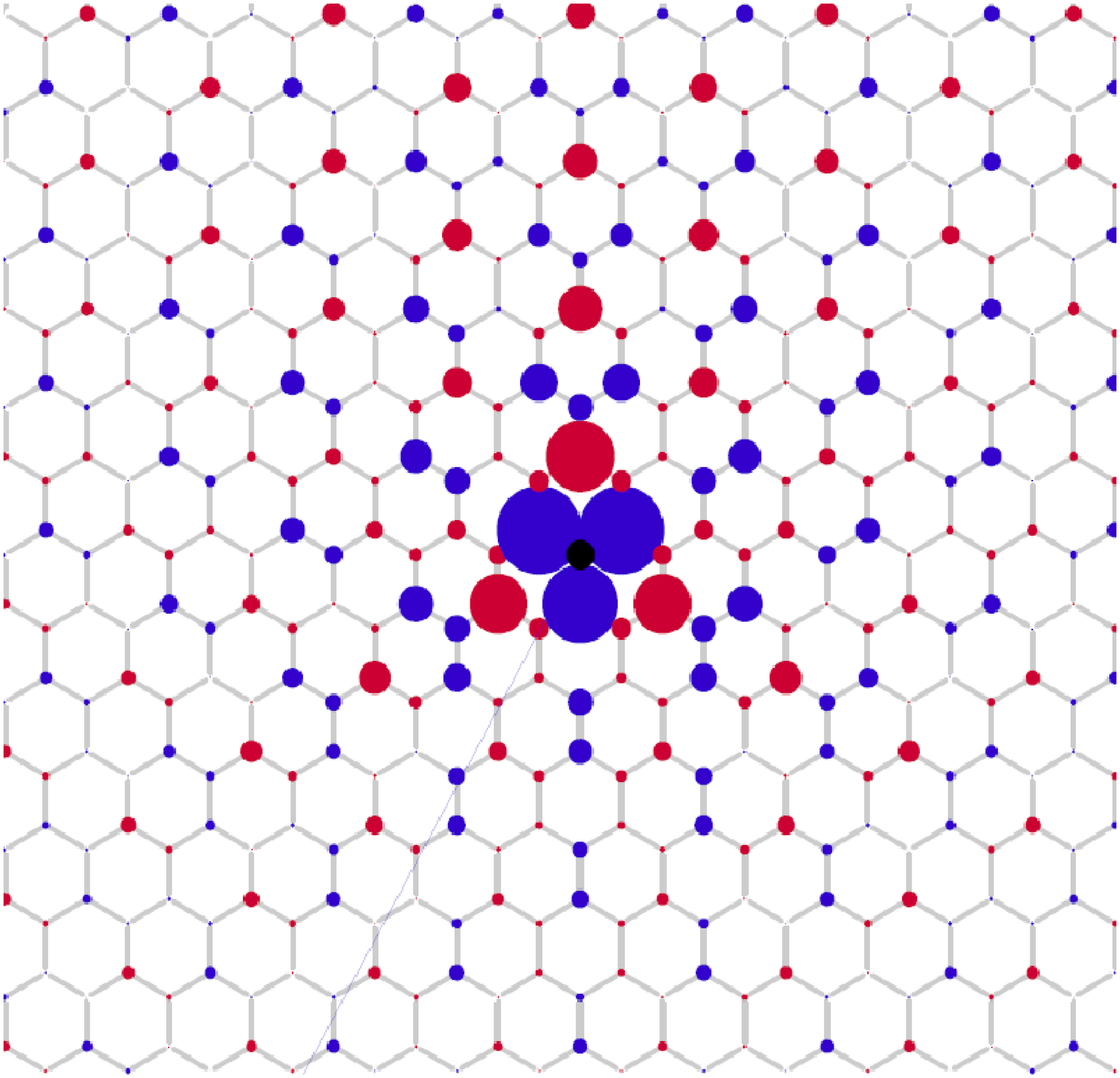}%
    \label{fig:Eigenstate-Tlinha-2}%
  }%
  \caption{
    \coloronline
    Selected eigenstates in a graphene sheet with $80^2$ atoms containing a
    single impurity at the center (black dot). Only the region near the vacancy
    is shown.
    \subref{fig:Eigenstate-NearZero} The eigenstate with energy
    closest, but different, to zero..
    \subref{fig:Eigenstate-AtZero} The eigenstate with $E=0$.
    \subref{fig:Eigenstate-Tlinha-1} and
    \subref{fig:Eigenstate-Tlinha-2} show the presence of two
    quasi-localized eigenstates even with $t=0.2t$.
  }
  \label{fig:Eigenstate-Pics}
\end{figure}
%


\subsubsection{Numerical Results -- Single Vacancy
\label{subsubsec:Graphene-Disorder-Vacancies-Numerical}}

The first calculation is the numerical verification of the exact
analytical result for the localized state in \eqref{eq:psi_x_y}. For
that, we consider the tight-binding Hamiltonian \eqref{eq:Hamiltonian}
and calculate numerically, via exact diagonalization, the full spectrum and
eigenstates in the presence of
a single vacancy. For some typical results we turn our attention to
Fig.~\ref{fig:Eigenstate-Pics}. There we plot a real-space
representation of some selected wavefunctions. This has been done by drawing a
circle at each lattice site, whose radius is proportional to the wavefunction
amplitude at that site, and whose color (red/blue) reflects the sign (+/-) of
the amplitude at each site. Thus bigger circles mean higher amplitudes. 
In the first panel, \ref{fig:Eigenstate-NearZero}, we are showing the
eigenstate with lowest, yet non-zero, absolute energy. It is visible that the
wavefunction associated with such state spreads uniformly across the totality of
the system. like a plane wave. In the second panel,
\ref{fig:Eigenstate-AtZero}, we draw
the wavefunction of the state $E=0$, that corresponds to
\eqref{eq:psi_x_y}. The state is clearly decaying as the distance to
the central vacancy increases. In addition, the state exhibits the full $C_3$
point symmetry about the vacant site, just as expected. This picture provides a
snapshot of the lattice version\cite{Pereira:2005} of
\eqref{eq:psi_x_y}. Since only one vacancy was introduced, the state
shown in Fig.~\ref{fig:Eigenstate-AtZero} is the only zero mode
present.

When particle-hole symmetry is disturbed by a non-zero $t'$, we still find
states having this quasi-localized nature, where the wavefunction amplitude
is still quite concentrated about the vacancy. Two examples are shown in panels
\ref{fig:Eigenstate-AtZero} and
\ref{fig:Eigenstate-Tlinha-1}. They are two eigenstates with
neighboring energy calculated for the same system. An important difference
occurs here, in that, unlike the case $t'=0$ where only one localized state
appears, the particle-hole asymmetric case opens the possibility for more than
one of such states. 

\begin{figure}
  \centering
  \subfigure[][]{%
    \includegraphics*[width=\columnwidth]{%
      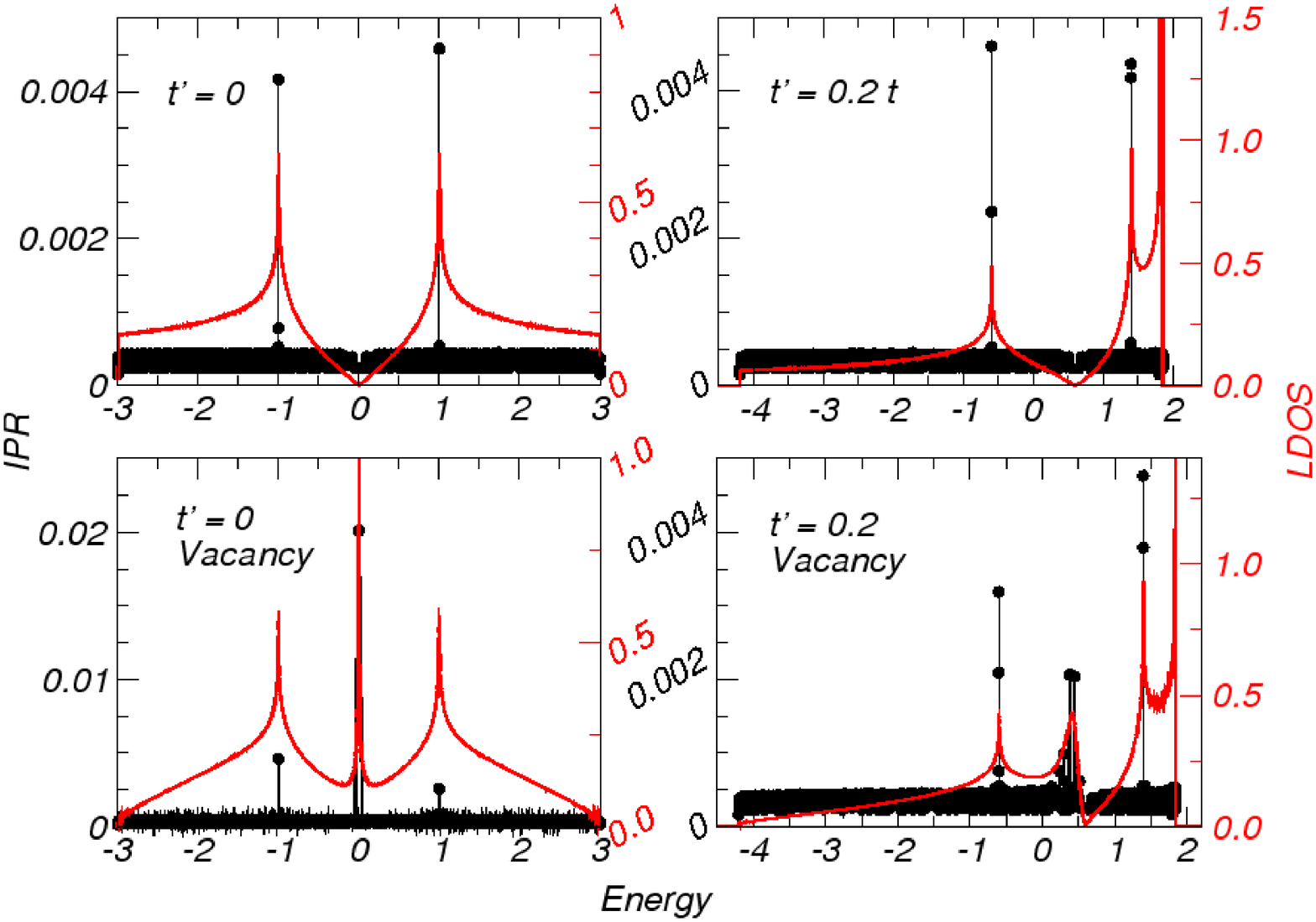}%
    \label{fig:PR-1}%
  }
  \subfigure[][]{%
    \includegraphics*[width=0.95\columnwidth]{%
      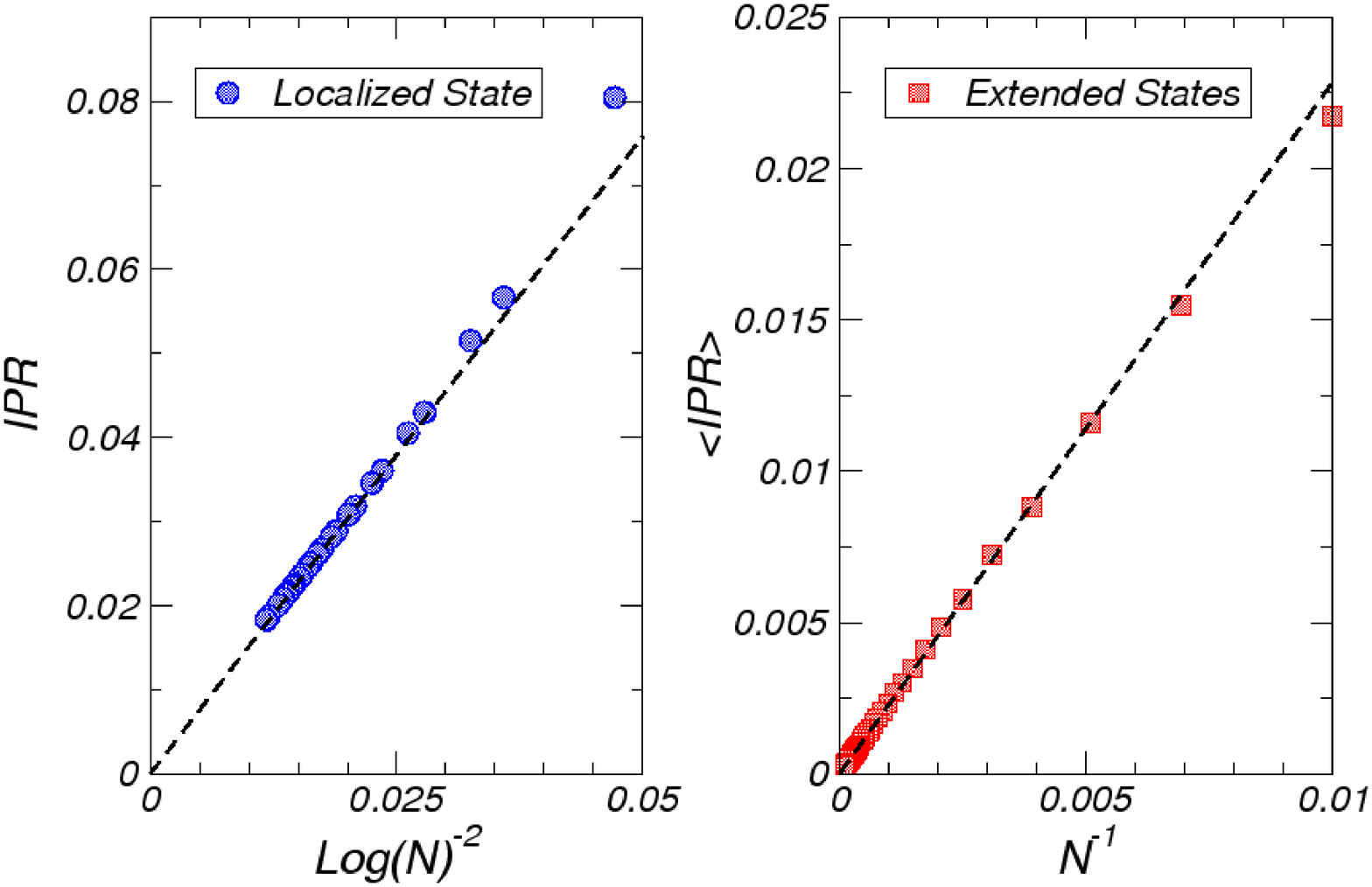}%
    \label{fig:PR-2}%
  }
  \caption{
    \coloronline
    \ac{IPR} and \ac{LDOS} calculated at one site closest to the vacancy.
    In panel \subref{fig:PR-1}, we have results for the \ac{IPR} with
    $t'=0$ and $t'\ne 0$ without any vacancy (top row), and with a single 
    vacancy (bottom) for comparison.
    In panel \subref{fig:PR-2} we show the dependence of the \ac{IPR}
    of the zero mode, $\mathcal P(E=0)$, with the system size $N$ (left),
    and also $\average{\mathcal P(E)}$ versus $N$ for the remainder (extended)
    states (right). Dashed lines are guides for the eye.
  }
  \label{fig:PR}
\end{figure}

This fact can be seen more transparently through the \acf{IPR} of the
eigenstates. With such purpose in mind, the \ac{IPR}
\begin{equation}
  \mathcal P(E_n) = \sum_i \abs{\Psi_n(r_i)}^{4}
  \nonumber 
\end{equation}
was calculated across the band in both the $t'=0$ and $t'\ne 0$ cases, with a
single central vacancy. Typical results are shown in Fig.~\ref{fig:PR}.
%
%
From \ref{fig:PR-1} we do confirm that, when $t'=0$, the presence of a
vacancy introduces a localized state at $E=0$, which is reflected both by the
enhanced \ac{IPR} there, and by the sharply peaked \ac{LDOS} calculated at the
vicinity of the vacancy site. Although not shown in this figure, the amplitude
of the peak in the \ac{LDOS} at $E=0$, $\rho_i(0)$, decays as the distance
between $\Ri$ and the vacancy increases, in total consistence with the
analytical picture. When next-nearest neighbor hopping is included, we also
confirm the appearance of states with a considerably enhanced \ac{IPR}. Not only
that, but, instead of one, we do observe a set of states with \ac{IPR} much
larger than the average for the remainder of the band. The \ac{LDOS} is also
enhanced near these energies, although the effect appears as a 
resonance on account of the finite \ac{DOS}, in contrast with the sharp peak in
the previous, particle-hole symmetric, case.

A more definite and quantitative analysis is provided by the results in the
subsequent panel (Fig.~\ref{fig:PR-2}). Here we present the dependence
of $\mathcal P(E)$ on the number of carbon atoms in the system, $N$. To
understand the differences, we recall that the \ac{IPR} for extended states
should scale as 
\begin{equation}
  \mathcal P(E) \sim \frac{1}{N}
  \,.
\end{equation}
But, for the zero mode (it should be obvious that when the term \emph{zero mode}
is employed, we are referring to the case with $t'=0$.), we face an interesting
circumstance. Remember that the wavefunction \eqref{eq:psi_x_y} is not
normalizable. So, strictly speaking, the state is not localized, and hence the
designation \emph{quasi-localized} that we have adopted above. The consequence
of this is that the normalization constant for $\Psi(x,y)$ depends on the system
size:
\begin{equation}
  \sum_i^N \abs{\Psi(x,y)}^2 \sim \log(\sqrt{N}) \sim \log(N)
  \,. 
\end{equation}
This, in turn has an effect on the \ac{IPR} because $\mathcal P(E)$ is defined
in terms of normalized wavefunctions:
\begin{equation}
  \mathcal P(0) = \frac{1}{\log(N)^2}
    \sum_i^N \abs{\Psi(x,y)}^4 \sim \frac{1}{\log(N)^2}
  \,. 
\end{equation}
This scaling of the \ac{IPR} with $N$ is precisely the one obtained numerically
in Fig.~\ref{fig:PR-2} (left) for the zero mode, and is just another
way of confirming the $1/r$ decay of this wavefunction.


\subsubsection{Numerical Results -- Finite Concentration of Vacancies}  

\begin{figure}
  \centering
  \subfigure[][]{\includegraphics*[width=0.85\columnwidth]{%
    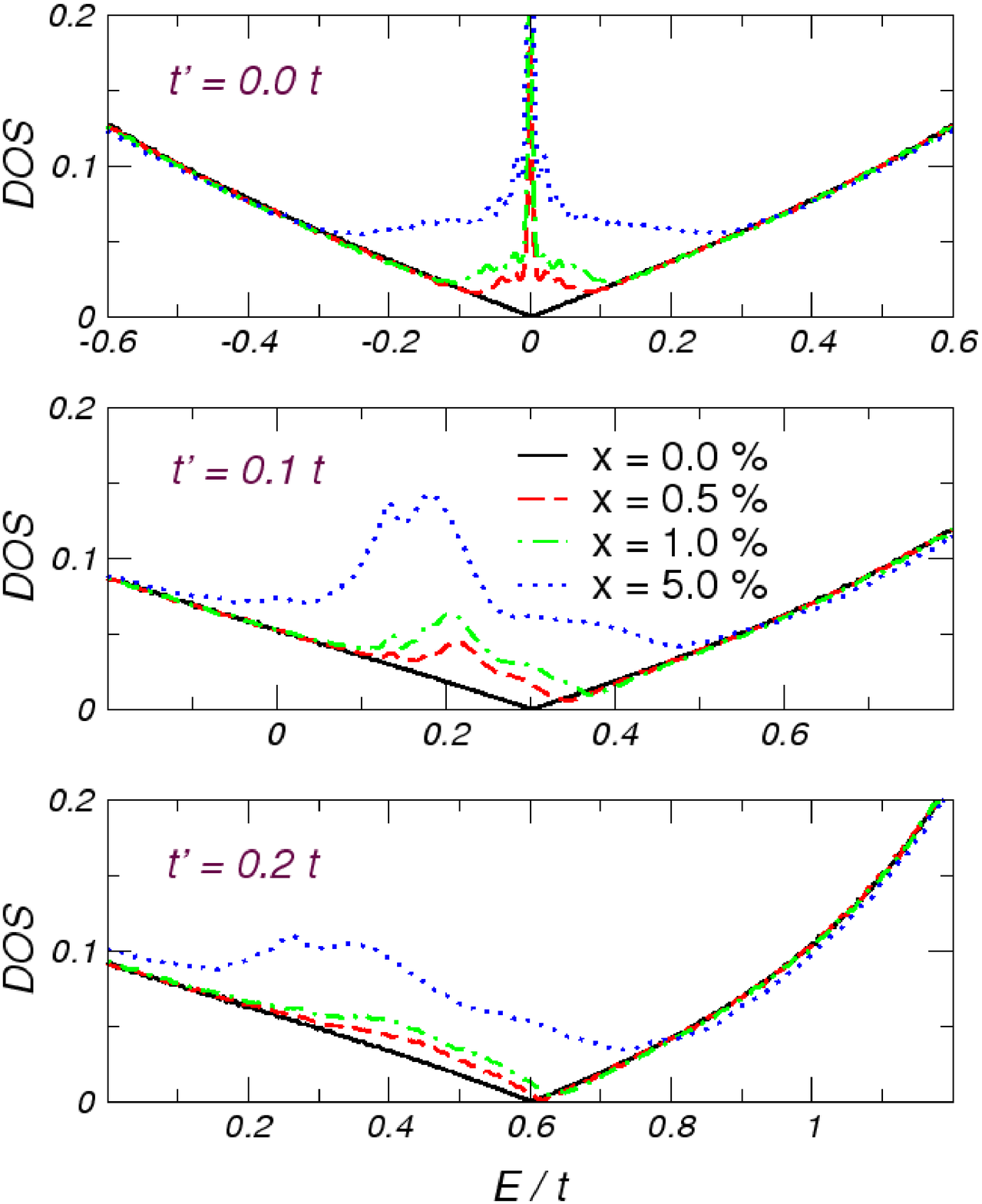}
    \label{fig:Dilution-LDOS}
  }
  \subfigure[][]{\includegraphics*[width=\columnwidth]{%
    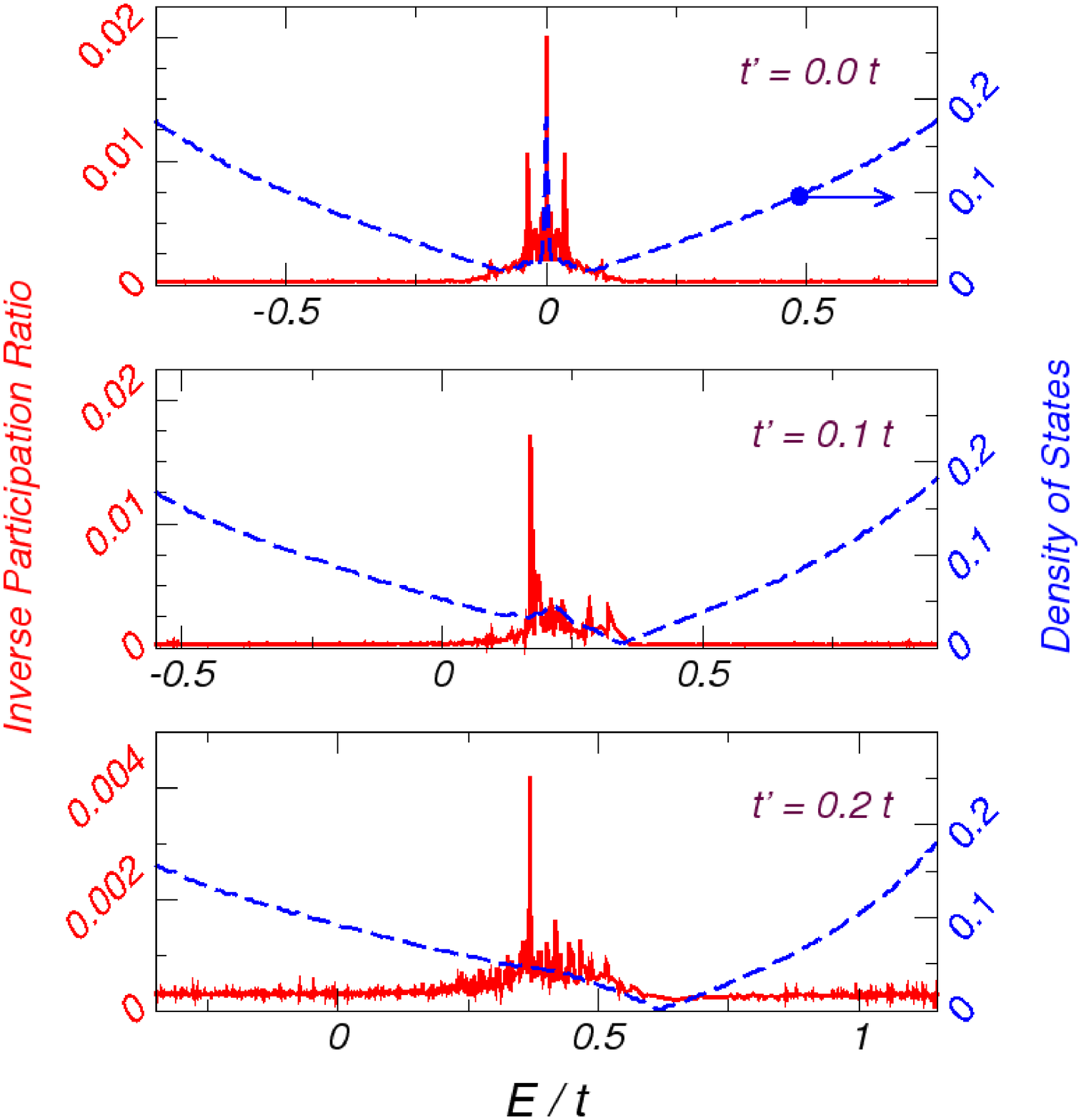}
    \label{fig:Dilution-PR}
  }
  \caption{
    \coloronline
    \ac{IPR} and \ac{DOS} for the diluted honeycomb lattice. 
    \subref{fig:Dilution-LDOS} \ac{DOS} for selected
    concentrations, $x$, and different values of $t'$. 
    \subref{fig:Dilution-PR} \ac{IPR} for selected values of $t'$ 
    using a concentration $x=0.5\,\%$. For comparison, the corresponding 
    \ac{DOS} is also plotted in each case. The concentration
    of vacancies is $x$, and only the vicinity of the Fermi level is shown.
  } 
  \label{fig:Dilution}
\end{figure}

Unlike the single vacancy case, the dilution of the honeycomb via the
introduction of a finite concentration of vacancies is not solvable using the
analytical expedients employed in ref.~\onlinecite{Pereira:2005}, and numerical
calculations become essential in this case. Our procedure consists in diluting
the honeycomb lattice with a constant concentration of vacancies, which we call
$x$ ($x=N_\text{vac}/N$). The diluted sites are chosen at random
and the global \ac{DOS}, averaged over several vacancy configurations, is
calculated afterwards. This is clearly a disordered problem, and we employ the
recursive method allowing us to obtain the \ac{DOS} for systems with $2000^2$
sites (which is already of the order of magnitude of the number of atoms in 
real mesoscopic samples of graphene studied experimentally).
Some results are summarized in Fig.~\ref{fig:Dilution}.
%
%
One of the effects of this disorder is, as always, the softening of the
van-Hove singularities (not shown). But the most significant changes occur in
the vicinity of the Fermi level (Fig.~\ref{fig:Dilution-LDOS}). In the
presence of electron-hole symmetry ($t'=0$), the inclusion of vacancies brings
an increase of spectral weight to the surroundings of the Dirac point, leading
to a \ac{DOS} whose behavior for $E\approx0$ mostly resembles the results
obtained elsewhere within \ac{CPA} \cite{Peres:2005a}. Indeed, for higher
dilutions, there is a flattening of the \ac{DOS} around the center of the band
just as in \ac{CPA}.
The most important feature, however, is the emergence of a sharp peak at the
Fermi level, superimposed upon the flat portion of the DOS (apart from the peak,
the DOS flattens out in this neighborhood as $x$ is increased past the $5\,\%$
shown here). 
The breaking of the particle-hole symmetry by a finite $t'$ results in the 
broadening of the peak at the Fermi energy, and the displacement of its 
position by an amount of the order of $t'$. All these effects take place 
close to the the Fermi energy. At higher energies, the only deviations from 
the \ac{DOS} of a clean system are the softening of the van~Hove singularities 
and the development of Lifshitz tails (not shown) at the  band edge, 
both induced by the increasing disorder caused by the random dilution. 
The onset of this high energy regime, where the profile of the \ac{DOS} is
essentially unperturbed by the presence of vacancies, is determined by 
$\epsilon \approx v_{\rm F} / l$, with $l\sim n_{\rm imp}^{-1/2}$ being
essentially the average distance between impurities.

To address the degree of localization for the states near the Fermi level, the
\ac{IPR} was calculated again, via exact diagonalization on smaller systems.
Results for different values of $t'$ are shown in
Fig.~\ref{fig:Dilution-PR} for random dilution at $0.5\,\%$. One
observes, first, that ${\mathcal P}_m \sim 3/N$ for 
all energies but the Fermi level neighborhood, as expected for states
extended up to the length scale of the system sizes used in the numerics. 
Secondly, the \ac{IPR} becomes significant exactly in the same energy 
range where the \ac{DOS} exhibits the vacancy-induced anomalies discussed
above. 
Clearly, the farther the system is driven from the particle-hole 
symmetric case, the weaker the localization effect, as illustrated 
by the results obtained with $t'=0.2\,t$. To this respect, it is 
worth mentioning that the magnitude of the strongest peaks in 
${\mathcal P}_m$ at $t'= 0$ and $t'=0.1\,t$ is equal to the magnitude 
of the \ac{IPR} calculated above for a single impurity problem. 
Such behavior indicates the existence of quasi-localized states at 
the center of the resonance, induced by the presence of the vacancies. 
For higher doping strengths, the enhancement of ${\mathcal P}_m$ is weaker 
in the regions where the \ac{DOS} becomes flat. This explains the qualitative
agreement between our results and the ones obtained within \ac{CPA} in that
region, since \ac{CPA} does not account for localization effects.

In summary, in  this section, we saw that a single vacancy
introduces a quasi-localized zero mode. Its presence is ensured by the
uncompensation between the number of orbitals in the two sublattices, and a
theorem from linear algebra. The presence of this mode translates in the
appearance of a peak in the \ac{LDOS} near the vacancy, and in an enhanced
\ac{IPR} for this state. When we go from one to a macroscopic number of
vacancies, we saw that both the peak and the enhancement of the \ac{IPR}
persist in the global \ac{DOS} at $E_F$ .


\subsection{Selective Dilution%
\label{subsec:SelectiveDilution}}

It is important to recall that the results of the previous section pertain to
lattices that were randomly diluted. During such process, we expect the
number of vacancies in sublattice $A$ to be equal to the number of
vacancies in sublattice $B$, on average. 
Strictly speaking, since our original lattices
are always chosen with $N_A = N_B$, the fluctuations on the degree of
uncompensation, $N_A - N_B$, should scale as $1/\sqrt{N}$ thus vanishing in the
thermodynamic limit. Because of this, in principle, we would expect the lattices
used above to be reasonably compensated. But the theorem in
\textsection~\ref{subsubsec:Graphene-Disorder-Vacancies-Theorem} only
guarantees the
presence of zero modes when the lattice is uncompensated.
It turns out that, notwithstanding our utilization of rather large system sizes,
such $\sqrt{N}$ fluctuations are still significant and the lattices were indeed
slightly uncompensated.

This clearly begs the clarification of the origin of the zero modes in the cases
with finite densities of vacancies. Do they appear only through these
fluctuations in the degree of sublattice compensation, or can we have zero modes
even with full compensation? To try to elucidate this we developed a controlled
approach to this issue, in the following. From now on we consider only the
particle-hole symmetric situation ($t'=0$).


\subsubsection{Complete uncompensation} 

%
\begin{figure}
  \centering
  \subfigure[][]{%
    \includegraphics*[width=0.8\columnwidth]{%
      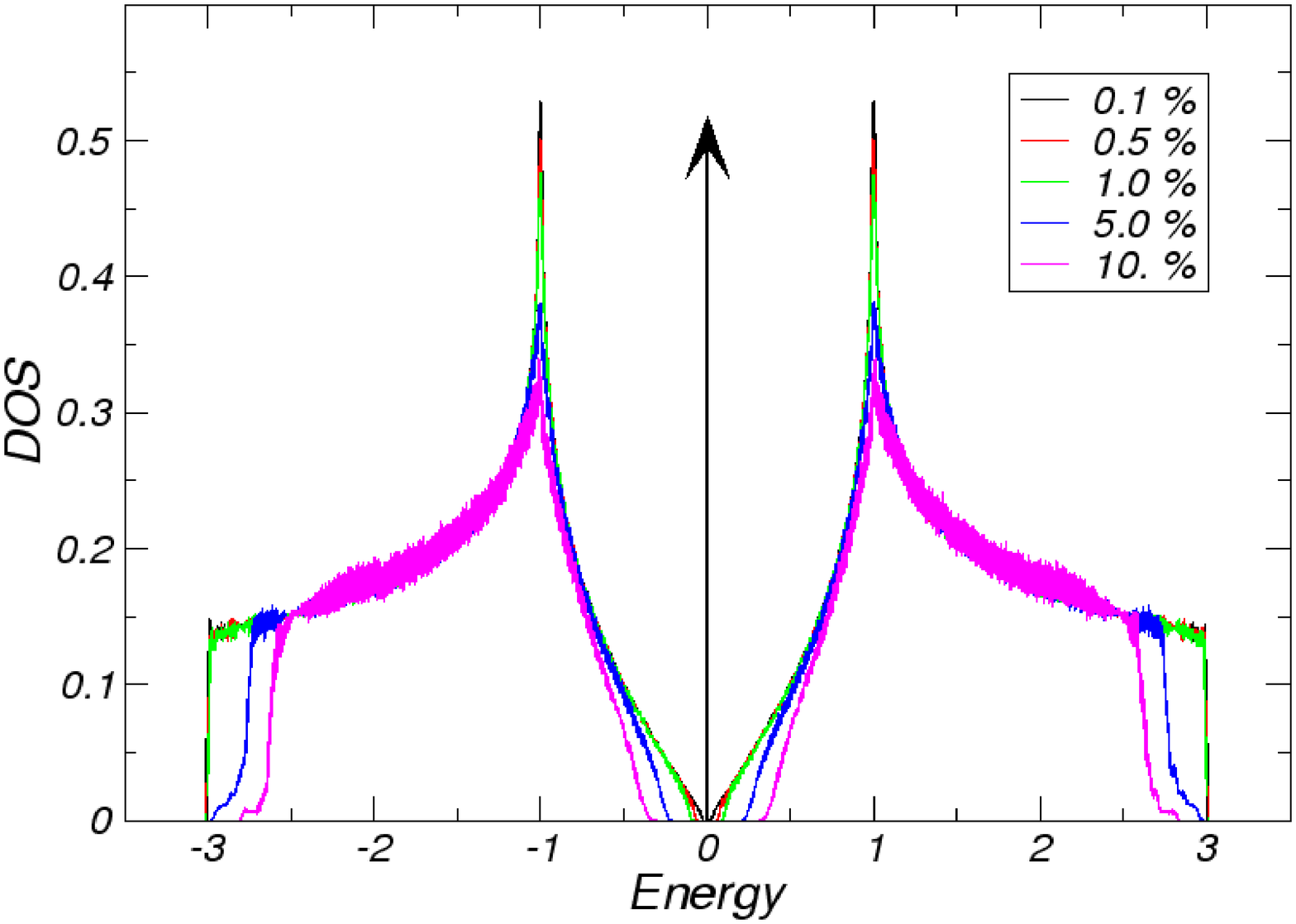}%
    \label{fig:SameSublattDilution-1}%
  }\\
  \subfigure[][]{%
    \includegraphics*[width=\columnwidth]{%
      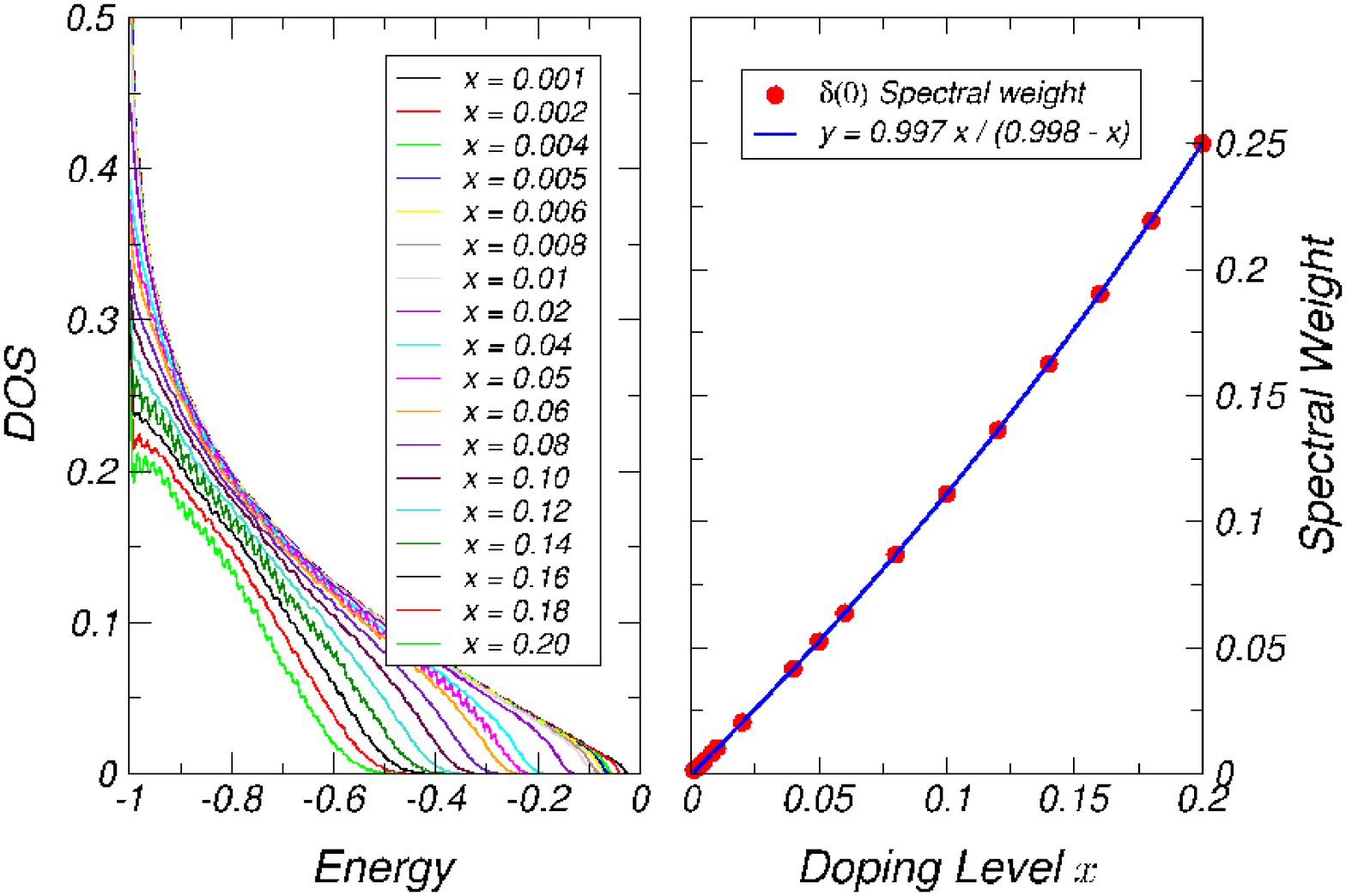}%
    \label{fig:SameSublattDilution-2}%
  }
  \caption{
    \coloronline
    Dilution of just one sublattice of the honeycomb.
    \subref{fig:SameSublattDilution-1} \ac{DOS} for different dilution
    strengths, diluting only sublattice $A$.
    \subref{fig:SameSublattDilution-2} On the left we show a detail of
    the \ac{DOS} and the evolution of the gap with vacancy concentrations. 
    On the right we plot the dependence of the missing spectral weight on
    the band ($=1-w_\delta$) with $x$ (circles). The continuous line is the
    best fit using $f(x) = a x /(b - x)$ to the data represented by the
    circles.     
    }
  \label{fig:SameSublattDilution}
\end{figure}
%

We have studied the \ac{DOS} for
systems in which only one of the sublattices was randomly diluted, with a finite
concentration of vacancies. 
In this case,  the system has  precisely a
number of zero modes that equals the number of vacancies. Starting from a clean
lattice with $N=N_A+N_B$ sites, the latter corresponds to $N_v = N x$. We should
thus expect a $\delta(E)$ peak contributing to the global \ac{DOS}, with an
associated spectral weight, $w_\delta$ that coincides with the fraction of zero
modes:
\begin{equation}
  w_\delta(x) = \frac{N x}{N (1-x)} = \frac{x}{1-x}
  \,.
  \label{eq:SpectralWeightDelta}
\end{equation}
Since the total spectral weight is normalized to 1, the spectral weight at $E=0$
has to be transferred from the states in the band. 
In Fig.~\ref{fig:SameSublattDilution} we show  what is happening. 
As seen in  \ref{fig:SameSublattDilution-1}, the selective dilution promotes
the appearance of a gap in the
\ac{DOS}, whose magnitude increases with the amount of dilution. At the center
of the gap we can only see an enormous peak (not visible in the range used)
staying precisely at $E=0$, corroborating our expectations regarding the
Dirac-delta in the \ac{DOS}. But since it appears exactly at $E=0$, we cannot
resolve numerically its associated spectral weight. To obtain such spectral
weight we calculated the spectral weight loss in the remainder of the band. 
The result and its variation with the amount of dilution, $x$, is
displayed in the right-most frame of
Fig.~\ref{fig:SameSublattDilution-2}. A non-linear fit to the data
reveals that the dependence expected from
\eqref{eq:SpectralWeightDelta} is indeed verified by the accord
between the fitted curve in Fig.~\ref{fig:SameSublattDilution-2} and
eq.~\eqref{eq:SpectralWeightDelta}.

%
\begin{figure}
  \centering
  \includegraphics*[width=0.9\columnwidth]{%
    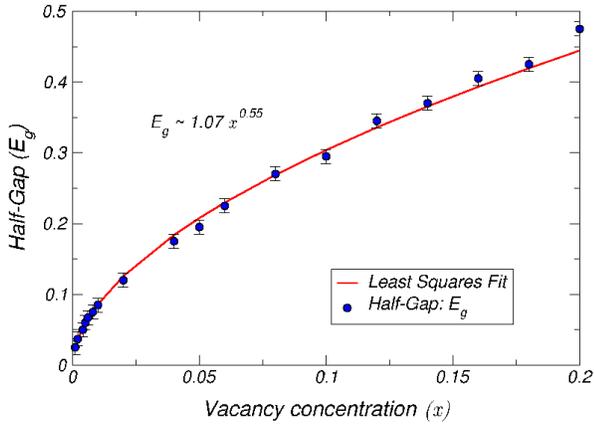}
  \caption{
    \coloronline
    The gap estimated from the numerical curves in
    Fig.~\ref{fig:SameSublattDilution} is plotted against the vacancy
    concentration, $x$. The continuous line is a least squares fit to 
    $f(x) = ax^b$. 
  }
  \label{fig:Dilution-Gap}
\end{figure}
%
As Fig.~\ref{fig:SameSublattDilution} shows, the spectral weight is
transferred almost entirely from the low energy region near $E_F$ and from the
high energy regions at the band edges. This depletion near $E=0$ introduces the
gap, $2 E_g$. A gap implies the existence of a new energy scale in the problem.
Since the hopping $t$ is the only energy scale in the Hamiltonian, such new
scale has to come from the concentration of vacancies. By dimensional analysis,
such scale is dictated essentially by the average distance between vacancies
($l$)  
\begin{equation}
  \epsilon \sim \frac{v_F}{l} \sim n_{\text{vacancies}}^{1/2} \sim \sqrt{x}
  \qquad (\hbar = 1)
  \,.  
\end{equation}
When the the magnitude of the gap found numerically is plotted against $x$ we
arrive at the curve of Fig.~\ref{fig:Dilution-Gap}. 
The least squares fit shown superimposed onto the numerical circles leaves
confirms  this assumption, and we arrive at a quite interesting
situation, of having a half-filled, particle-hole symmetric and gapped system,
with a finite concentration of (presumably quasi-localized) zero modes at the
mid-gap point.


\subsubsection{Controlled uncompensation} 

\begin{figure}
  \centering
  \includegraphics*[width=\columnwidth]{%
    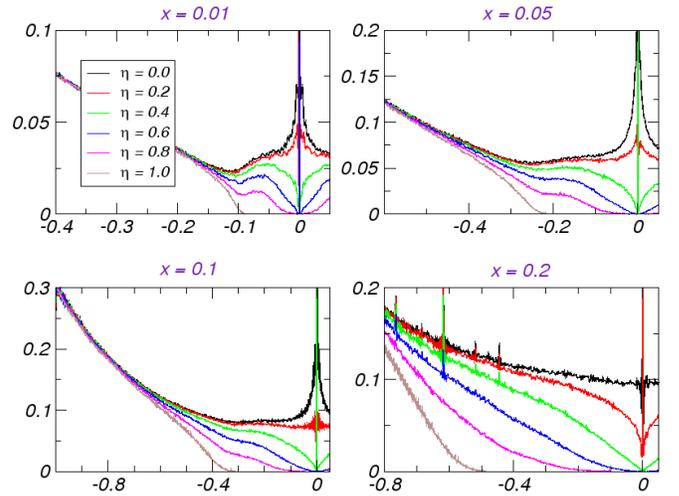}
  \caption{
    \coloronline
    \ac{DOS} for the honeycomb lattice using the controlled selective 
    dilution discussed in the text, calculated for different concentrations 
    of vacancies, $x$, and several degrees of uncompensation, $\eta$. 
    Only the low energy region close to the Dirac point is shown.
  }
  \label{fig:Dilution-Controled}
\end{figure}

We now turn to a more controlled
approach to the dilution and uncompensation. For that we  introduce an
additional parameter, $\eta$, that  measures the degree of uncompensation. As
before, we want to study finite concentrations of vacancies. This is determined
by $x$ in such a way that the number of vacancies in a lattice with $N$ sites
will be $N_v = Nx$. But now, the number of vacancies in each sublattice is
determined by
\begin{align}
  N_v^A &= \frac{1}{2} N x (1 + \eta) \nonumber \\
  N_v^B &= \frac{1}{2} N x (1 - \eta)
  \label{eq:Dilution-Controled}
  \,,
\end{align}
with $0\le\eta\le 1$. Therefore, the parameter $\eta$ permits an interpolation
between completely uncompensated dilution ($\eta = 1$), and totally compensated
dilution ($\eta = 0$). Let us look directly at the results for the \ac{DOS},
calculated at different $x$ and $\eta$, and plotted in
Fig.~\ref{fig:Dilution-Controled}.
%
%

At any concentration $x$ the following sequence of events unfolds as $\eta$
decreases from 1 to 0: 
(i) There is a perfectly defined gap in the limit $\eta=1.0$ discussed above;
(ii) for $\eta\lesssim 1$ a small hump develops at the same energy scale of the
previous gap; 
(iii) although the gap seems to disappear, it is clearly visible
that when $\eta\lesssim 1$, the \ac{DOS} decays to zero after the hump in a
pseudo-gap like manner, and is zero at $E=0$; 
(iv) decreasing further $\eta$
towards complete compensation (say, for $\eta=0.6,\,0.4$), this behavior
persists, being visible that the \ac{DOS} drops to zero at $E=0$; 
(v) closer to full compensation ($\eta=0.1$) the \ac{DOS} seems to display an
upward inflection near $E=0$, and apparently does not drop to zero. 
Unfortunately, we are unable to resolve this region numerically with the desired
accuracy. For instance, at higher dilutions $x=0.2$ we can still see the curve
of $\eta=0.2$ dropping to zero near $E=0$. 

Naturally that, for all the cases with $\eta\ne 0$, the existence of $N_v^A -
N_v^B$ zero modes is guaranteed. As before, we inspected this by calculating the
missing spectral weight in the bands, and confirmed that it does agree with the
fraction of uncompensated vacancies. Hence, the picture emerging from these
results seems to suggest that, although the gap disappears for $\eta < 1$, the
\ac{DOS} still drops to zero at $E=0$, and might drop in a singular way as
$\eta$ approaches zero. If we separate the contributions of the zero modes to
the global \ac{DOS} from the contribution of the other states, the consequence
of this would be that, in a compensated lattice ($\eta=0$), the \ac{DOS}
associated with the other states would seem to diverge as $E\to0$, but would be
zero precisely at $E=0$. Stated in another way, coming from high energies, we
would see a decreasing \ac{DOS} up to some typical energy $\epsilon \sim
\sqrt{x}$, at which point it would turn upwards. At very small energies the
\ac{DOS} would seem to be diverging but, at some point arbitrarily close to
$E=0$, it would drop precipitously down to zero. 
Unfortunately, at the moment the numerical calculations are not so accurate as
to allow the confirmation or dismissal of such possibility. In fact, the peaks
for $\eta=0.0$ are of the same magnitude of the ones found when the dilution is
completely random across the two sublattices
(Fig.~\ref{fig:Dilution-LDOS}). So, although the evidence is compelling
towards the affirmative, these results are still inconclusive as to whether the
zero modes disappear in a perfectly compensated diluted lattice or not.


\subsection{Local Impurities \label{subsec:Disorder-LocalImpurities}}

Vacancies are local scatterers in the unitary limit. A vacancy can be thought as
an extreme case of a local potential, $U$, when $U\to\infty$. In this context we
investigated the intermediate case characterized by a finite local potential.
The Hamiltonian in this case changes from the pure tight
binding in \eqref{eq:Hamiltonian} to 
\begin{equation}
  H = U \sum_p  c^\dagger_p c_p
      - t \sum_{\langle i,j \rangle} c^\dagger_i c_j
      - t' \sum_{\langle\langle i,j \rangle\rangle} c^\dagger_i c_j + \text{
h.c. }
  \label{eq:Hamiltonian-finiteU}
  \,. 
\end{equation}
The first term represents the local potential of magnitude $U$ at the impurity
sites $p$. These impurity sites belong to the underlying honeycomb lattice but
their space distribution is random. The concentration of
impurities, $x = N_i / N$, is kept constant and we consider only the case with
$t^\prime=0$ in the sequel. 

Physically the model summarized in the Hamiltonian of
eq.~\eqref{eq:Hamiltonian-finiteU} could describe the situation in
which some of the carbon atoms are substituted by a different species. Another
realistic circumstance has to do with the fact that a real graphene sheet is
expected to have some molecules from the environment adsorbed onto its
surface\cite{Schedin:2007}. 
Consequently, even if the honeycomb lattice of the
carbon atoms is not disrupted with foreign atoms, the presence of adsorbed
particles can certainly induce a local potential at the sites where they couple
to the carbon lattice. 

Much of the details of this model can be understood from the local
environment around a single impurity, in which case exact results and closed
formulas are obtainable within a T-matrix approach\cite{Elliott:1974}. Hence,
we start by analyzing the single impurity problem in the honeycomb lattice,
taking into account the full electronic dispersion and calculating the exact
local Green's functions, which allow the identification of the main
spectral changes introduced by the scattering potential.
%
%
\begin{figure}
  \centering
  \includegraphics*[width=\columnwidth]{%
    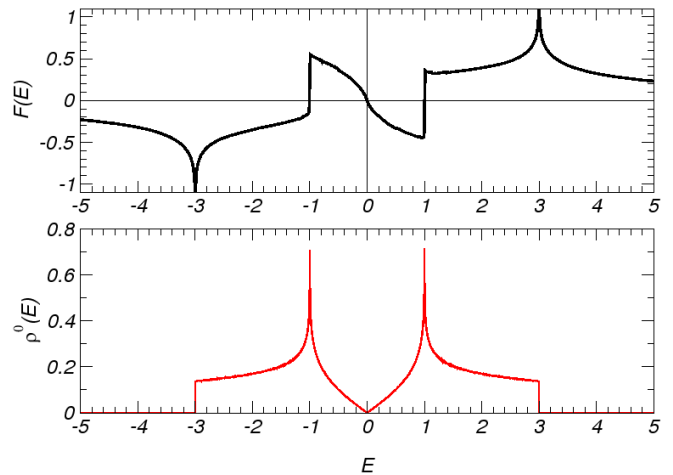}%
  \caption{
    \coloronline
    Real part of \eqref{eq:KramersKronig} (top) obtained from the homogeneous 
    \ac{DOS} (bottom) through the Kramers-Kronig relations.
  }
  \label{fig:KramersKronig}
\end{figure}
%
Within T-matrix, the $2\times 2$
electron Green's function is written as 
\begin{equation}
  \bG_{r,r'} = \bG^0_{r,r'} + \sum_{s,s'}
      \bG^0_{r,s} \bT_{s,s'} \bG^0_{s',r'}
  \,.
\end{equation}
In the Dyson-like expansion above $\bG^0$ is the non-interacting Green's
function
whose matrix elements are denoted by $[\bG^0]_{r,r'}^{\alpha,\beta}$
(sub/superscripts refer to position/sublattice), and the
T-matrix, $\bT$, is formally defined in terms of the scattering potential,
$\bV$, by\cite{Elliott:1974,Doniach:1999}
\begin{equation}
  \bT(E) = \frac{ \bm{V} }{1-\bG^0 \bV}
  \,.
\end{equation}
Taking 
$\bV^{\alpha\beta}_{r,r'} = U
\delta_{r,r'}\delta_{r,0}\delta_{\alpha,\beta}\delta_{\alpha,0}$ for a
potential localized only on site $r=0$ of sublattice A, the local Green's
function on that site reads
\begin{equation}
  G^{AA}_{00} = \frac{[G^0]^{AA}_{00}}{1 - U [G^0]^{AA}_{00}}
  \,.
  \label{eq:GF}
\end{equation}
The function $[G^0]^{AA}_{00}$ is simply related to the density of states
\emph{per carbon atom} in the absence of impurity, $\rho^0(E)$, through 
\begin{equation}
  [G^0(E)]^{AA}_{00} = F(E) - i \pi \rho^0(E)
  \,.
  \label{eq:KramersKronig}
\end{equation}
%
%
\begin{figure}
  \centering
  \includegraphics*[width=\columnwidth]{%
    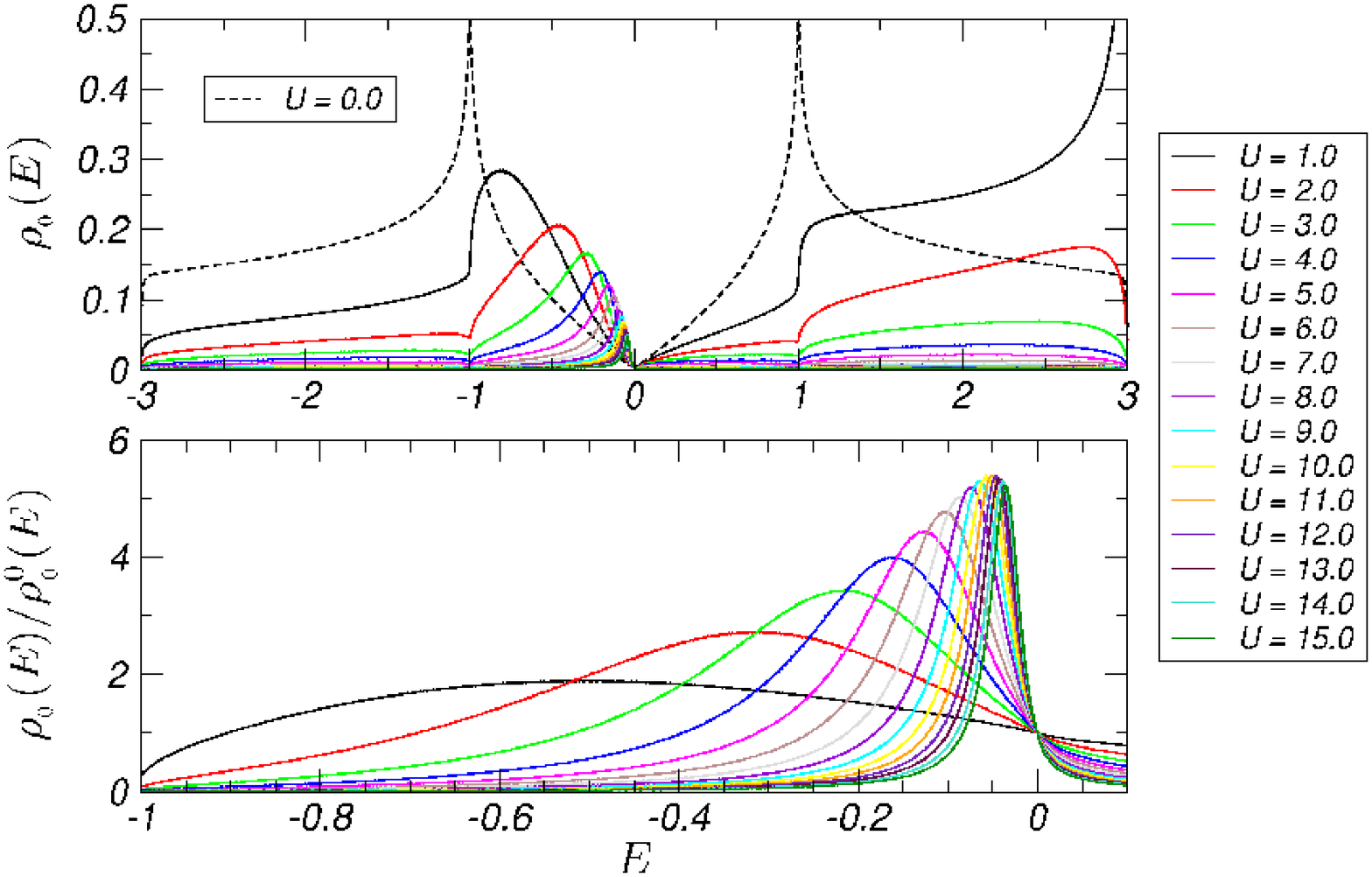}%
  \caption{
    \coloronline
    (Top) \ac{LDOS} at the impurity site ($\rho_0(E)$) for different strengths
    of the scattering potential, indicated in the legend.
    (Bottom) The same data divided by the free \ac{DOS} ($\rho^0(E)$).
  }
  \label{fig:LDOS-At-Local-Imp-Site}
\end{figure}
%
The knowledge of $\rho^0(E)$ suffices for the determination of $F(E)$ on
account of the analytical properties of $\bG^0(E)$ and the Kramers-Kronig
relations. Moreover, any new poles of the exact Green's function can
come only from the denominator in \eqref{eq:GF}, and are determined by the
condition
\begin{equation}
  1 - U F(E) = 0
  \,.
  \label{eq:ResonanceCondition}
\end{equation}
Should this condition be satisfied for $E$ within the branch cut of $\bG^0$,
the new poles will signal the existence of resonant states in the band, and
bound states of the local potential otherwise.
Since $\rho^0(E)$ is known exactly\cite{Hanisch:1995} (cfr.
Fig.~\ref{fig:DOS-Comparison}), so can be $\bG^0(E)$ through
eq.~\eqref{eq:KramersKronig}. The function $F(E)$ is shown in
Fig.~\ref{fig:KramersKronig}. The profile of this function and the
condition above, allows two immediate conclusions without further
calculation: (i) the presence of the
local potential induces bound states beyond the band continuum and (ii) a
resonance appears at low energies beyond a certain threshold, $U_\text{res}$,
with energy of opposite sign with respect to
the scattering potential, and which moves toward $E=0$ as $U\to\infty$. 

\begin{figure}
  \centering
  \subfigure[][]{%
    \includegraphics*[width=0.48\columnwidth]{%
      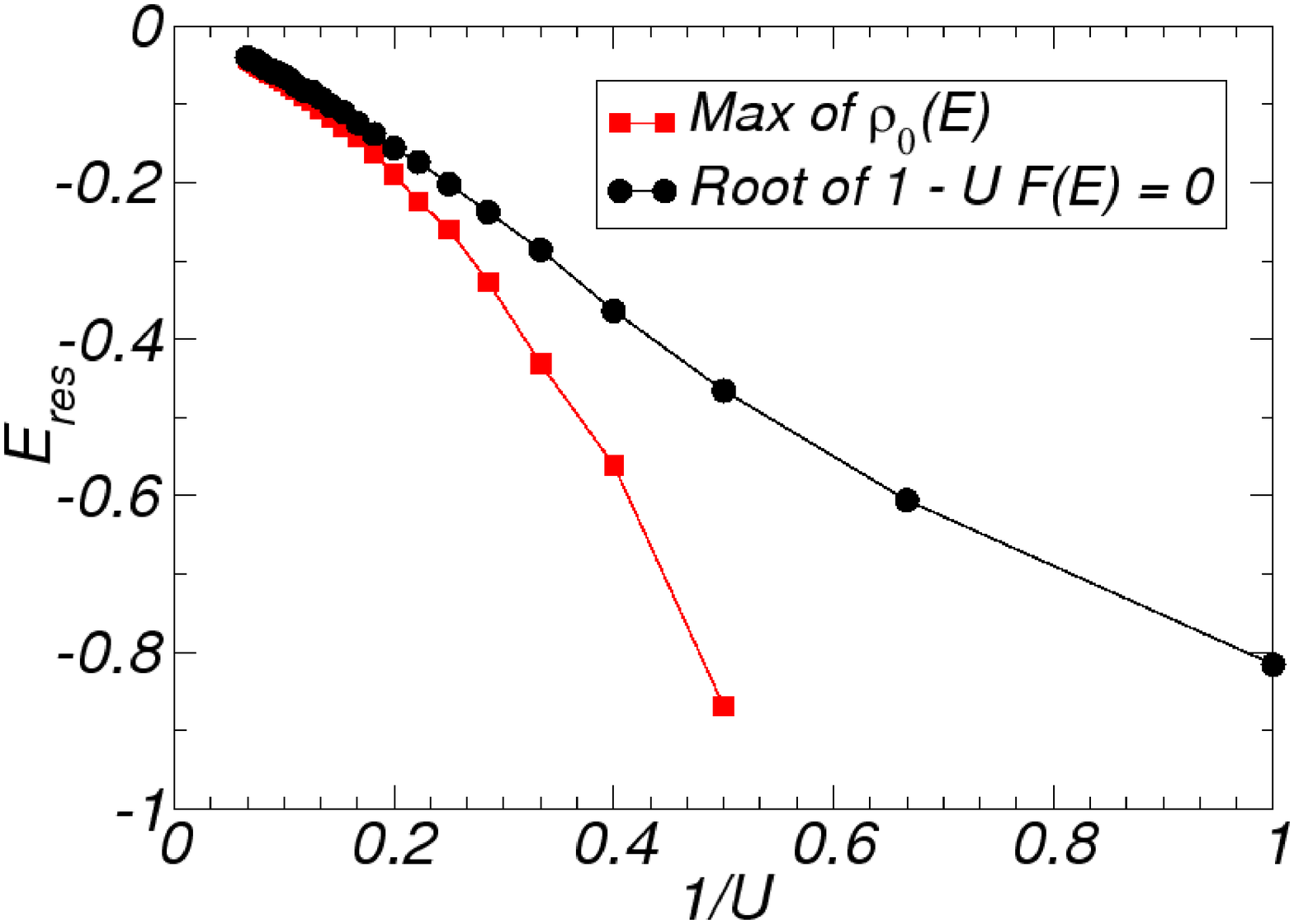}%
    \label{fig:Binnary-Eres}%
  }
  \subfigure[][]{%
    \includegraphics*[width=0.48\columnwidth]{%
      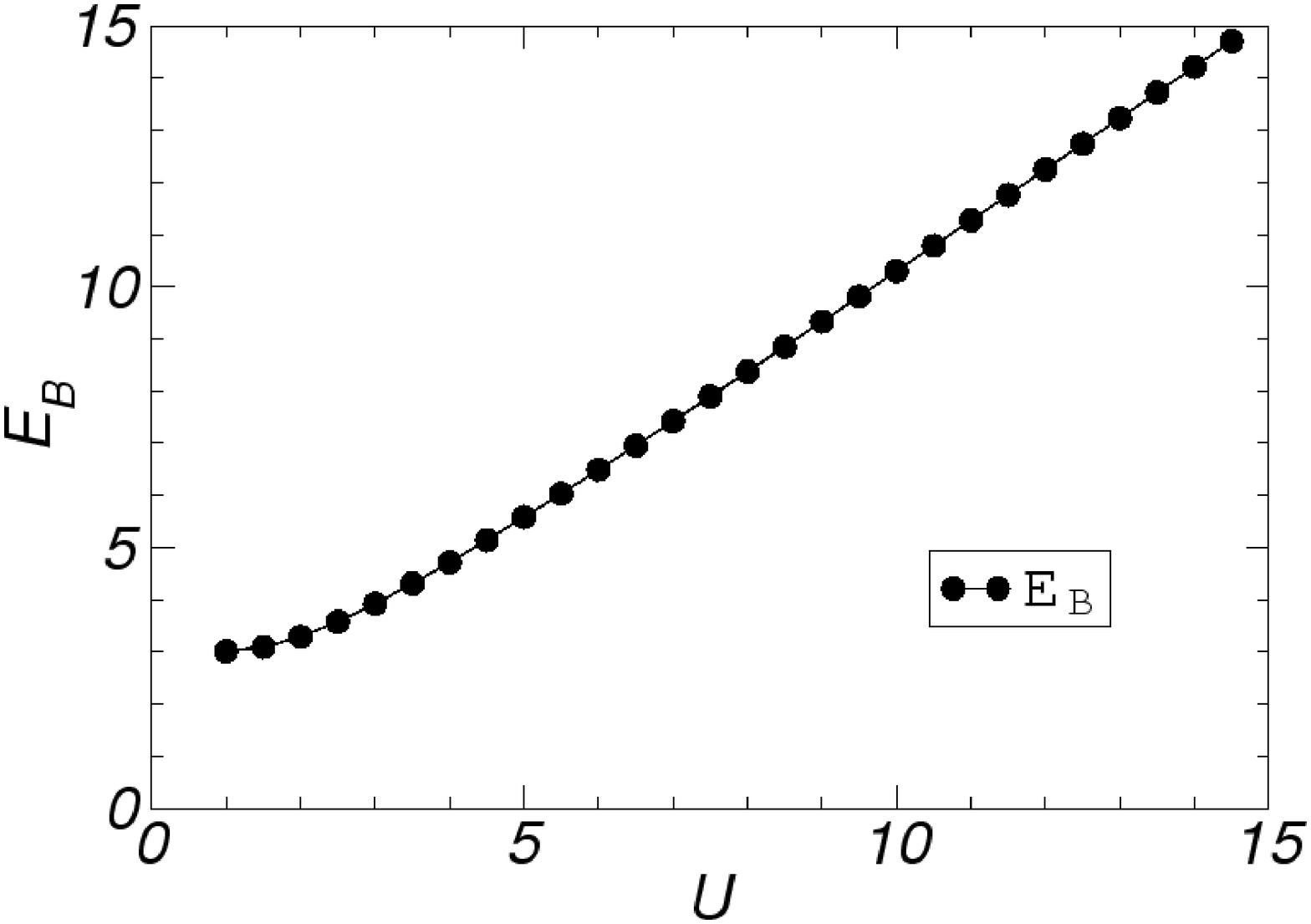}%
    \label{fig:Binnary-Ebound}%
  }
  \caption{
    \coloronline
    \subref{fig:Binnary-Eres} Position of the maximum in the \ac{LDOS} compared
    with the roots of \eqref{eq:ResonanceCondition}.
    \subref{fig:Binnary-Ebound} Energy of the bound state.
  }
  \label{fig:Eres-and-Ebound}
\end{figure}

The latter characteristic is certainly more interesting and we explore it a
little further. To that extent notice that from \eqref{eq:GF} and
\eqref{eq:KramersKronig} follows the interacting \ac{LDOS} at the impurity site:
\begin{equation}
  \rho_0(E) = \frac{\rho^0(E)}
    {\bigl[(1-U F(E)\bigr]^2 + \bigl[\pi U \rho^0(E)\bigr]^2}
  \,.
  \label{eq:LDOS-AtImpSite}
\end{equation}
This quantity was calculated using the results of Fig.~\ref{fig:KramersKronig}
and \eqref{eq:LDOS-AtImpSite} and, at the same time, using the recursive
method that we have been using so far. The two do coincide, just as expected
since the solution of the single impurity problem is exact, and, on the other
hand, the recursive method is exact for the particular case
of the \ac{LDOS}\cite{Haydock:1972a}. The \ac{LDOS} at the site of the impurity
is shown in the top frame of Fig.~\ref{fig:LDOS-At-Local-Imp-Site} for several
values of $U$. The bottom frame shows the same data divided by the
non-interacting \ac{DOS}, which amounts to replacing
$\rho^0(E)$ by unity in the numerator of eq.~\eqref{eq:LDOS-AtImpSite}. The
resonance alluded above is visible in both panels through the marked
enhancement of the \ac{LDOS} in the vicinity of the Dirac point. The position
of the maximum in $\rho_0(E)$ differs slightly from the roots of
eq.~\eqref{eq:ResonanceCondition} due to the modulation introduced by
$\rho^0(E)$ in \eqref{eq:LDOS-AtImpSite}. This effect is shown in detail in
Fig.~\ref{fig:Binnary-Eres} where the two values are explicitly compared. In
addition, the \ac{LDOS} also exhibits the Dirac-delta peak associated with the
bound state (not shown in the figure), whose energy is plotted in
Fig.~\ref{fig:Binnary-Ebound} as a function of $U$.
%
%
%
It is worth mentioning that analytical expressions can be obtained for the
resonant condition \eqref{eq:ResonanceCondition} using the low energy Dirac
approximation to the electronic dispersion\cite{Peres:2005a,Skrypnyk:2006}. 

\begin{figure}
  \centering
  \subfigure[][]{%
    \includegraphics*[width=\columnwidth]{%
      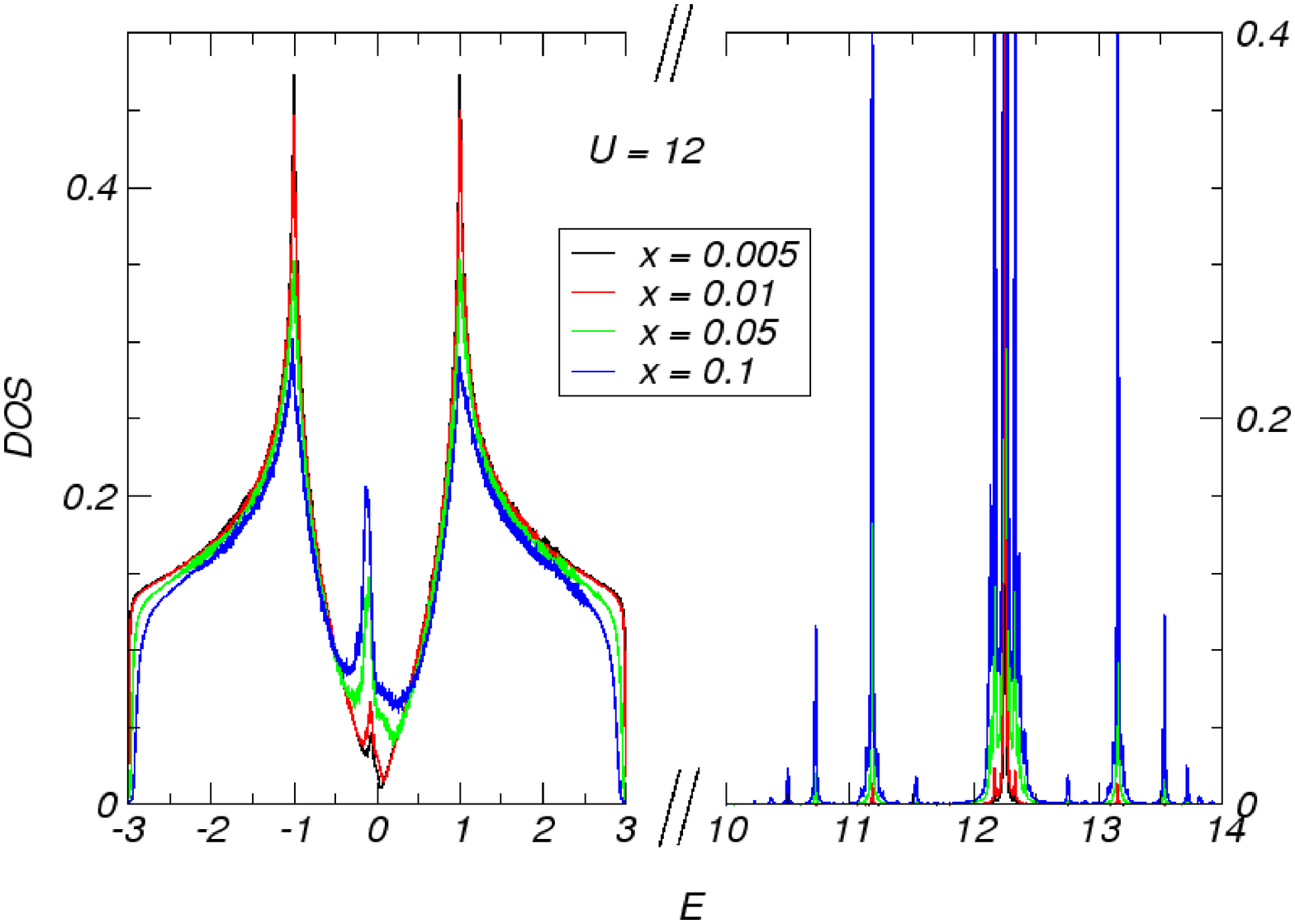}%
    \label{fig:Binnary-Ea-12}%
  }
  \subfigure[][]{%
    \includegraphics*[width=\columnwidth]{%
      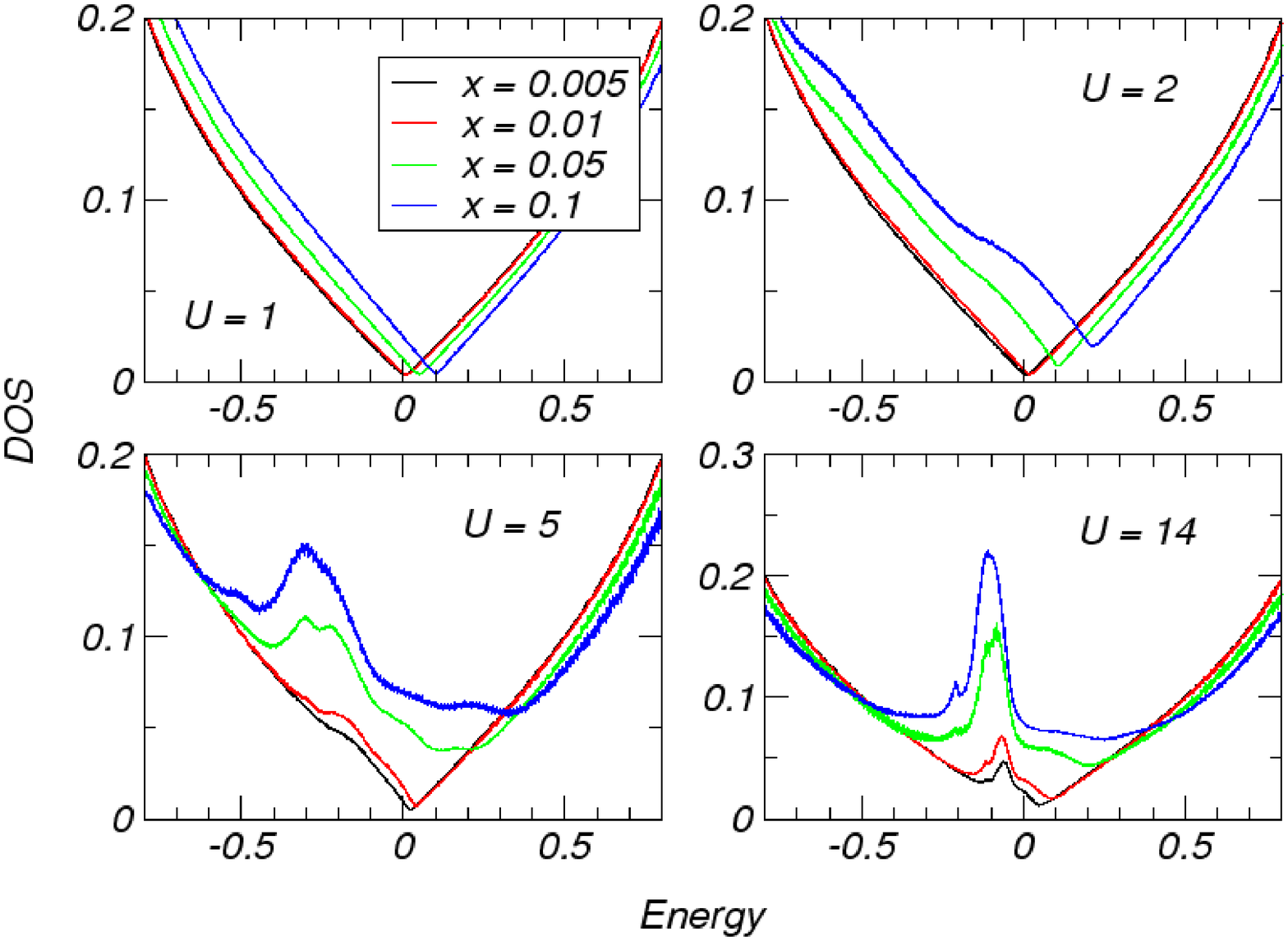}%
    \label{fig:Binnary-DOS-vs-Ea-and-x}%
  }
  \caption{
    \coloronline
    DOS of the honeycomb lattice with a finite density of local impurities.
    \subref{fig:Binnary-Ea-12} shows the \ac{DOS} for $U=12t$ and different
    concentrations of impurities (notice the truncation in the horizontal 
    axis).
    \subref{fig:Binnary-DOS-vs-Ea-and-x} shows a detail of the low energy
    region for different $U$ and $x$ as noted in the different graphs. 
  }
  \label{fig:Binnary}
\end{figure}

Returning now to our initial goal of populating the lattice with a finite
concentration of local impurities, we expect the main features of the above
analysis to hold to a large extent. But new features should also emerge from the
possibility of multiple scattering and interference effects in a multi-impurity
environment. Although some of
these effects can be captured within standard approximations to impurity
problems\cite{Doniach:1999}, we choose to present the exact numerical results
obtained with the recursion technique. Examples of such calculations are shown
in Fig.~\ref{fig:Binnary}, where the global \ac{DOS} averaged over
several configurations of disorder is shown for different potential strengths
and concentrations.
%
%

\begin{figure}
  \centering
  \subfigure[][]{%
    \includegraphics*[width=0.9\columnwidth]{%
      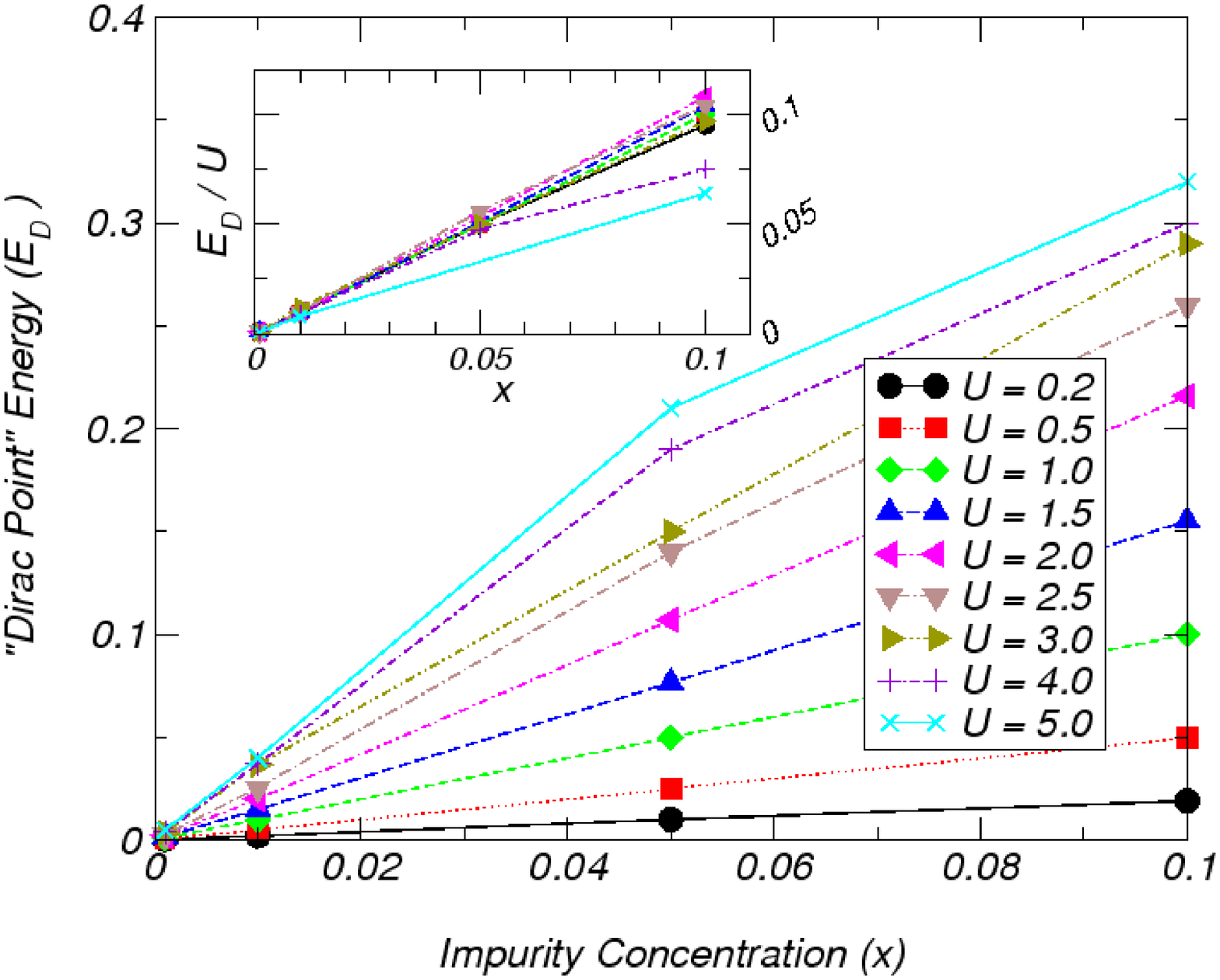}%
    \label{fig:Binnary-DiracPt-vs-x-and-U}%
  }
  \hfill%
  \subfigure[][]{%
  \includegraphics*[width=0.9\columnwidth]{%
    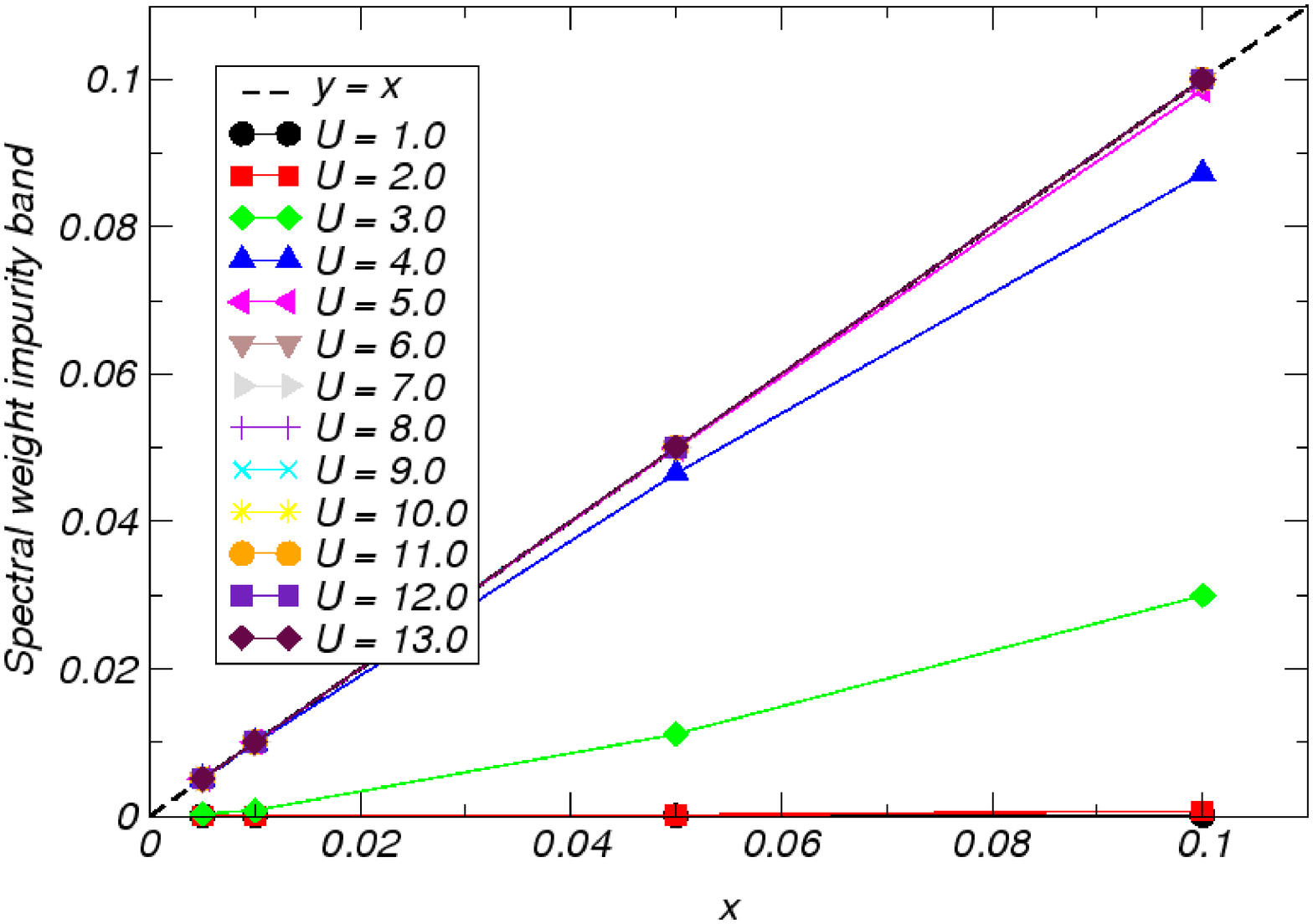}%
    \label{fig:Binnary-SW-ImpBand}%
  }%
  \caption{
    \coloronline
    \subref{fig:Binnary-DiracPt-vs-x-and-U}
    Variation of the \emph{``Dirac point''} energy $E_D$ with impurity 
    concentration and strength. The inset shows $E_D/EU$ as a function of $U$, 
    in which the curves with $U\le 3$ roughly collapse onto each other.
    \subref{fig:Binnary-SW-ImpBand}
    Spectral weight transfer to the impurity band in the presence of local
    impurities. The values shown correspond to integration of the global
    \ac{DOS} beyond $E=4 t$ (above the main band edge, cfr.
    Fig.~\ref{fig:Binnary}).
  }
  \label{fig:Binnary-2}
\end{figure}

The presence of the local term clearly destroys
the particle-hole symmetry, leading to the asymmetric curves in the figure. As
Fig.~\ref{fig:Binnary-Ea-12} makes clear, among the features seen locally for
a single impurity (Fig.~\ref{fig:LDOS-At-Local-Imp-Site}), the ones that carry
to the global \ac{DOS} of the thermodynamic system with a finite
concentration of impurities are the resonant enhancement of the \ac{DOS} in
the vicinity of the Dirac point, and the high energy features that dominate
beyond the band edge, and are associated with the impurity states. One verifies
that a finite concentration, $x$, generates a sort of impurity band at scales of
the order of $U$, in accordance with the results in
Fig.~\ref{fig:Binnary-Ebound}. 
This impurity band has an interesting splitted structure as can be seen in
the figure and is completely detached from the main band for $U\gtrsim4 t$.
In Fig.~\ref{fig:Binnary-DOS-vs-Ea-and-x}, we amplify the low energy region and
display what happens as $U$ and $x$ vary. At small $x$ and $U$ the 
the \ac{DOS} changes only through a simple translation of
the band with the concomitant shift in the Dirac point, $E_D$. This rigid shift
of the band at low disorder is simply a consequence of the rigid band
theorem\cite{Kittel:QTS}: it states that the form of
the \ac{DOS} in an alloy system does not change with alloying, other than via a
simple translation as given by first-order perturbation theory. In our case 
the magnitude of this shift is given by
\begin{equation}
  \Delta E = \Bigl\langle U \sum_p  c^\dagger_p c_p \Bigr\rangle
           \simeq x\, U
  \,,
  \label{eq:Binary-Band-Shift}            
\end{equation}
where an average over disorder is implied.
We can confirm that the exact numerical results satisfy quantitatively this
expectation by inspection of the data in
Fig.~\ref{fig:Binnary-DiracPt-vs-x-and-U}. There we plot the position of
the minimum in the \ac{DOS}, $E_D$, for several $U$ and $x$, being evident
that, for the concentrations analyzed, the relation \eqref{eq:Binary-Band-Shift}
is quite accurately satisfied up to $U\simeq3$.
%
%
For local potentials higher than $U\simeq 3$ the rigid shift of $E_D$ breaks
down and, in fact, the position of $E_D$ becomes slightly ill defined. We point
out that concurrently with the shift of $E_D$ (and the band), there is a marked
increase in the \ac{DOS} at $E_D$, unlike the single impurity case
(cfr. Fig.~\ref{fig:LDOS-At-Local-Imp-Site}). This, again, is expected and
appears already in approximate methods like the \ac{CPA}
approximation\cite{Peres:2005a}. Nonetheless, whereas $E_D$ shifts linearly for
moderate potential strengths, the position of the resonance does not vary
significantly with concentration, and is
only enhanced with an increasing number of impurities
(Fig.~\ref{fig:Binnary-DOS-vs-Ea-and-x}).

Another noteworthy aspect of this model has to do with the impurity band that
emerges at high energies. Besides the effects just described, a
change in the concentration of impurities implies a concomitant
redistribution of spectral weight between the main band and the impurity
band. This is plainly shown in Fig.~\ref{fig:Binnary-SW-ImpBand} which
displays the spectral weight in the impurity band against the concentration of
impurities. This spectral weight is calculated by integrating the \ac{DOS} in
the region $[4t,\infty[$. As the figure shows, for $U\gtrsim 5t$ the spectral
weight of the impurity band saturates at the value $x$, signaling the detachment
of the impurity states from the main band. For those cases the spectral weight
coincides with the concentration $x$. It certainly had to be so because with
increasing $U$ the 
impurity band drifts to higher energies, eventually disappearing from the
 problem in the unitary limit. As discussed previously in section
\ref{subsec:Disorder-Vacancies}, the spectral weight of the main band is
decreased by precisely $x$, in the presence of a concentration of vacancies
of $x$. This is totally consistent with the fact that the local impurity
interpolates between the clean case and the vacancy limit. 

Finally, is also clear how the vacancy limit ($U\to\infty$) emerges from the
data in Fig.~\ref{fig:Binnary} as the resonance approaches $E=0$ and
becomes more sharply defined. At the same time, the impurity band is displaced
toward higher and higher energies, eventually projecting out of the problem in
the vacancy limit.

  
\subsection{Non-Diagonal Impurities
\label{subsec:Disorder-SubstImpurities}}

%
\begin{figure}
  \centering
  \subfigure[][]{%
    \includegraphics*[width=\columnwidth]{%
      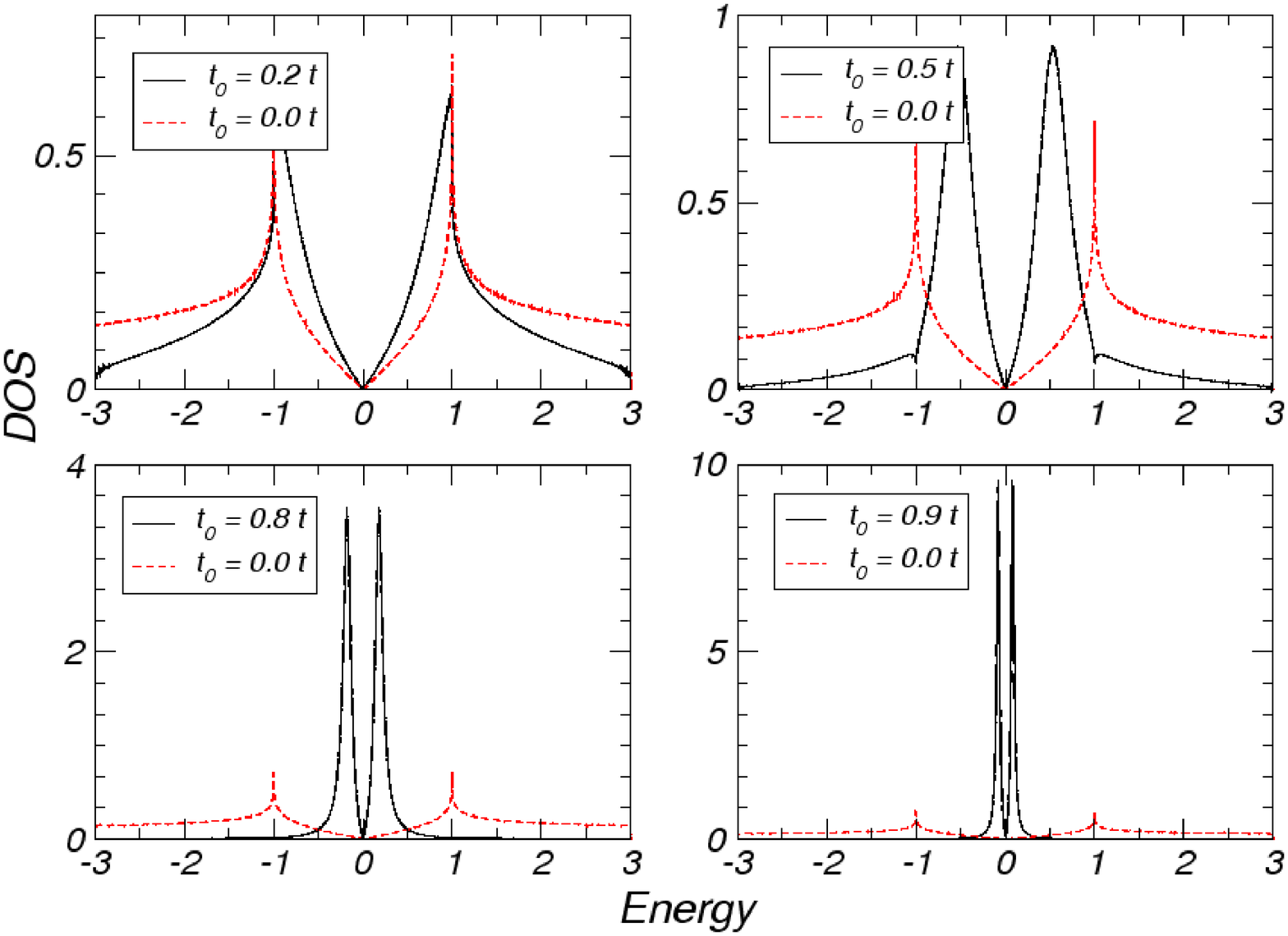}%
    \label{fig:Hopping-Pert-LDOS1}%
  }
  \subfigure[][]{%
    \includegraphics*[width=\columnwidth]{%
      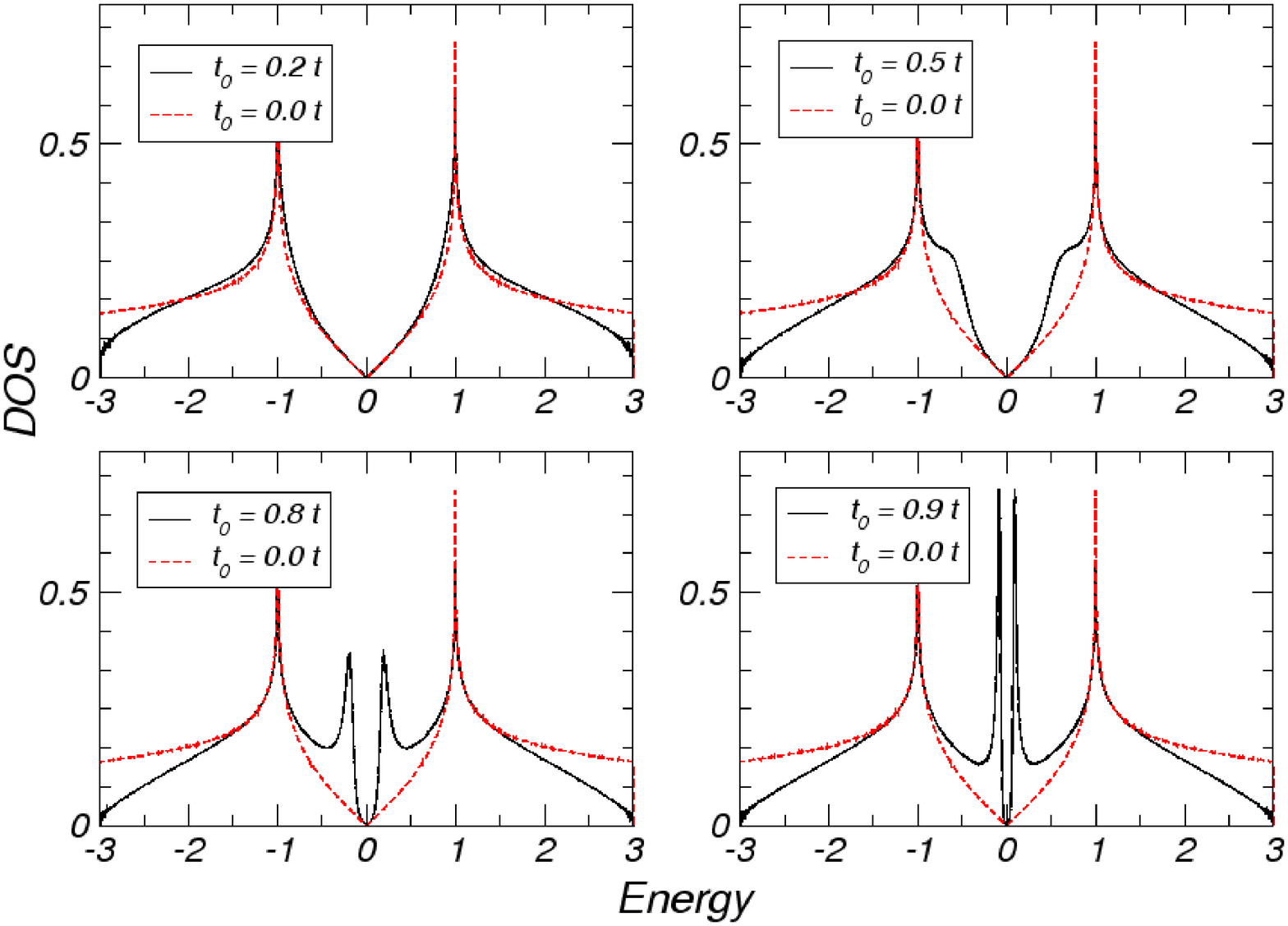}%
    \label{fig:Hopping-Pert-LDOS2}%
  }
  \caption{
    \coloronline
    Effect of a single substitutional impurity in the \ac{LDOS}. In panel
    \subref{fig:Hopping-Pert-LDOS1} we plot the \ac{LDOS} calculated 
    at the site of the impurity for the four different values of $t_0$ 
    indicated in each frame. In \subref{fig:Hopping-Pert-LDOS2} the 
    situation is identical but the \ac{LDOS} is calculated at the nearest 
    neighboring site of the impurity. 
  }
  \label{fig:Hopping-Pert-LDOS}
\end{figure}
%

Another effect expected with the inclusion of a substitutional impurity in the
graphene lattice is the modification of the hoppings between the new atom and
the neighboring carbons. This happens because the host and substituting atoms
have different radii, because the nature of the orbitals involved in the
conduction band is different, or, most likely, a combination of both. Customary
impurities in carbon allotropes are nitrogen, working as a donor, and boron,
working as an acceptor \cite{Kaiser:1959}. In fact, the selective inclusion of
nitrogen and/or boron impurities in carbon nanotubes is a current practice in
the hope to tune the nanotubes' electronic response
\cite{Droppa:2004,Ciraci:2004,Nevidomskyy:2003}. 

In general the study of a perturbation in the hopping is much less studied in
problems with impurities than the case of diagonal, on-site, perturbations. In
the context of our investigations, the perturbation in the hopping can, again,
be interpreted as an interpolation between a vacancy and an impurity. To be more
precise, let us introduce the relevant Hamiltonian: 
\begin{equation}
  H = - t \sum_{i, \delta} c^\dagger_i c_{i+\delta} 
      + t_0 \sum_{p, \delta}\null^\prime c^\dagger_p c_{p+\delta}
      + \text{h.c.}
  \label{eq:Hamiltonian-HoppingPert}
  \,.
\end{equation}
In this case, only nearest neighbor hopping is considered. Without the second
term, $H$ above is the Hamiltonian for pure graphene. The last sum is restricted
to the impurity sites, $p$, and $t_0$ represents a perturbation in the hopping
amplitude to its neighbors. Is plain to see that, when $t_0 = t$, all the
impurity sites turn into vacancies since the hopping thereto is zero. As a
result of that, this model provides another type of interpolation between pure
graphene and diluted graphene. An important difference is that this model can be
disordered when the impurities are placed at random, without breaking
particle-hole symmetry, and, in this sense, is qualitatively much different from
the case of local disorder discussed in the preceding section. 

We first look at the \ac{LDOS} in Fig.~\ref{fig:Hopping-Pert-LDOS}, which
contains typical results for
the local \ac{DOS} near the impurity, and at the impurity site itself.
Irrespective of whether the \ac{LDOS} is calculated at or near the impurity, the
resulting curves display a strong resonance in the low energy region, no bound
states are formed and the curves are symmetrical with respect to the origin. 
As $t_0$ increases from zero, two simultaneous modifications in these resonances
take place. 
The first is that they are clearly enhanced as $t_0$ approaches $t$. The
second is its shift in the direction of the Dirac point, in such a way that,
when $t_0 = 0.9t$, the peak is already very close to $E=0$. With regard to this
last point, we systematically investigated the variation of the peak position in
the \ac{LDOS} at the impurity site with the value of $t_0$. 
%
\begin{figure}
  \centering
  \includegraphics*[width=0.8\columnwidth]{%
    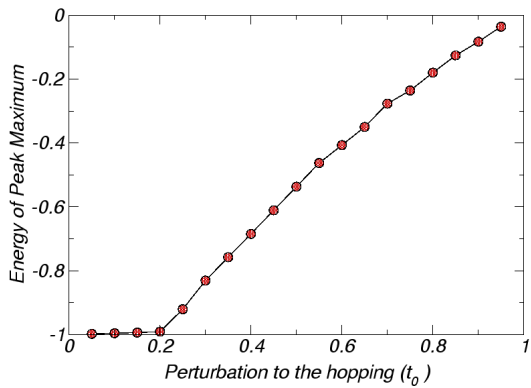}
  \caption{
    \coloronline
    Variation of the energy corresponding to the peak in the \ac{LDOS} with 
    the magnitude of $t_0$. The \ac{LDOS} in question is the \ac{LDOS} 
    calculated at the impurity site.
  }
  \label{fig:Hopping-Pert-Peak}
\end{figure}
%
This dependence, which can be seen in Fig.~\ref{fig:Hopping-Pert-Peak},
is approximately linear and, for $t_0 \gtrsim 0.6$, is reasonably well
approximated by the linear function $\epsilon_\text{max} \simeq t - t_0$. The
apparent saturation for smaller $t_0$ is due to the proximity to the van~Hove
singularity. The study of a single substitutional impurity has been also
undertaken in ref.~\onlinecite{Peres:2007}, with identical results.

The double-peak structure close to the Dirac point can be qualitatively
understood from the results regarding a vacancy. Suppose that one
completely severs the hopping between a given atomic orbital and its immediate
neighbors (i.e: set $t_0=t$). In this case we are left with an isolated orbital
with energy $E=0$ and a vacancy in the honeycomb lattice, which we know also has
a zero energy mode. If now $t_0$ is changed slightly, it will cause the
hybridization of the two zero energy modes with the  consequent splitting of
the energy level, and hence the double peaked structure of the \ac{LDOS} close
to the Dirac point.

\begin{figure}
  \centering
  \includegraphics*[width=\columnwidth]{%
    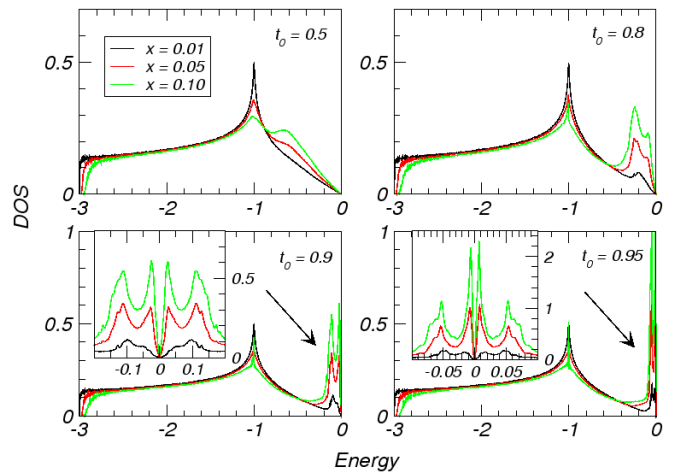}
  \caption{
    \coloronline
    The \ac{DOS} corresponding to the model Hamiltonian 
    \eqref{eq:Hamiltonian-HoppingPert}, with a finite density of
    impurities. The three panels correspond to different values of the 
    perturbing hopping ($t_0=0.5$, $0.8$, $0.9$ and $0.95$), and within each
    panel the three curves were obtained at different concentrations 
    ($x=0.01$, $0.05$ and $0.1$). The inset of the bottom panels is a 
    magnification of the region near $E=0$.
  }
  \label{fig:Hopping-Pert-DOS}
\end{figure}

When we go from one impurity
to a finite density of impurities, $x$, we obtain a
measurable influence in the thermodynamic limit. Our method in this case,
consists in placing impurities at random positions in the lattice, keeping their
concentration constant. The global \ac{DOS}, averaged over several realizations
of disorder, is presented in Fig.~\ref{fig:Hopping-Pert-DOS}. 
%
%
For intermediate values of $t_0$, the perturbation in the hopping induces a
resonance appearing at roughly the same energies as the ones found in
Fig.~\ref{fig:Hopping-Pert-Peak}. The resonance is enhanced at higher
concentrations of impurities, and becomes more sharply defined as $t_0\to t$.
Interestingly, as can be seen in the last panels of
Fig.~\ref{fig:Hopping-Pert-DOS} and its inset, the resonant peak splits at
higher perturbations. This splitting depends on the concentration of impurities
being more pronounced for larger concentrations and is a new feature
introduced by the finite number of impurities. As happened already in the case
of local impurities, the exact numerical results have qualitative and
quantitative features that could not be anticipated from calculations with a
single impurity within the usual approximation methods. We would also like to
point out the fact that, from inspection of the above figures, the \ac{DOS}
remains zero at $E=0$, notwithstanding the sharp resonances in its vicinity.
Since this model of disorder interpolates between clean graphene and graphene
with vacancies, we are led to a situation similar to the one encountered in
sec.~\ref{subsec:SelectiveDilution} for uncompensated vacancies. As before,
it seems that, as the vacancy limit is approached, the \ac{DOS} remains zero at
the Fermi energy, despite diverging arbitrarily close to this point, and so the
question of the \ac{DOS} exactly at $E=0$ for vacancies lingers. Furthermore,
unlike what happens with local impurities, there is no impurity band nor any
high energy features appearing as $t_0\to t$: the action is all on the low
energy regions. Strictly speaking, in the limit $t_0=t$, the impurity sites
become isolated from the carbon network. Hence those sites have to be
removed from the Hilbert space for a meaningful physical description of the
vacancy case as the limit $t_0\to t$ (for local impurities the removal of the 
impurity sites is akin to the drift of the impurity band to infinity, carrying
the spectral weight associated with the number of the impurities, which projects
out of the problem).

Before closing, just a comment on the physical origin of this perturbation. In
effect, the presence of a substitutional impurity like $N$ or $B$ will
introduce, simultaneously, a perturbation in the hopping, and in the local
energy. However, it is more or less clear from the discussions in the previous
section that the clearest resonances near $E_F$ occur when the local potential,
$U$, is moderate or high, which is not the case for boron or nitrogen
substituents.
Hence, the perturbation in the hopping should perhaps be more significant in
dictating the changes in the low-energy electronic structure in the real
physical system.


\section{Conclusions \label{sec:Conclusions}}

In this paper we have studied the influence of local disorder in the
electronic structure of graphene, within the tight-binding approximation of
eq.~\eqref{eq:Hamiltonian}. We focused on vacancies in an otherwise perfect
graphene plane and the not so extreme cases of local (diagonal) impurities and
substitutional (non-diagonal, or both) impurities. In all cases we saw that
disorder brings dramatic alterations of the spectrum in the vicinity of the
Fermi level. This is highly significant since many of the peculiar
physical properties of graphene stem from the vanishing of the \ac{DOS} at the
Dirac point.

In the case of vacancies, the \ac{DOS} features a strong
divergence at and close to $E=0$, which is associated with the formation of
quasi-localized states decaying as $\sim 1/r$ around the vacancies, which
remain even in the presence of next-nearest neighbor hopping.
Rather interesting is the particular case of lattices with uncompensated
vacancies, in which case we found the appearance of a gap at low
energies proportional to the concentration $x$, and the coexistence of
localized zero modes in the middle of this gap. For the extreme limit of
dilution among sites of a given sublattice only, we showed that the gap is
robust, and that a macroscopic number of quasi-localized zero modes dominates
the spectral density in the middle of the gap. 
Moreover, these zero modes are strictly non-dispersive as imposed by symmetry,
and give a contribution $x\,\delta(E)$ to the gapped \ac{DOS}.
This is very interesting, in
particular if one reasons in terms of magnetic instabilities and formation of
local magnetic moments. Such states might be at the origin of local magnetic
moments, which would explain the magnetism seen experimentally in the
irradiation experiments\cite{Esquinazi:2003}.

We showed how the vacancy case emerges as the limiting case of a local
impurity. In this case the exact calculation with a single impurity problem was
presented, taking into account the full dispersion of the honeycomb lattice.
The results of approximate methods such as \ac{CPA} were subsequently compared
with the exact numerical solution of the problem with finite concentrations of
impurities, and we identified the values of the parameters for which these
approximations qualitatively break down. The discussion of 
non-diagonal impurities provided yet another alternative view of the
interpolation between clean graphene and vacancies, with relevance for systems
with dopants that replace the host carbon atoms in the honeycomb lattice. One
important aspect of the results with a finite concentration of these impurities
regards the splitting of the low energy peaks 
(insets of Fig;~\ref{fig:Hopping-Pert-DOS}), which is not captured at a single
particle level. The effect has to do with situations in which substitutional 
impurities appear close to each other, causing interference and hybridization
effects that lead to the re-splitting of the low energy resonances.

Finally, the results provided for the \ac{DOS} and \ac{LDOS} are directly
testable in
real-life samples through scanning tunneling spectroscopy techniques and,
moreover, the effects on the global \ac{DOS} should reflect themselves in
the electric transport. For example, one might be able to distinguish whether
the
main effect of a substitutional impurity occurs through the modification of the
hopping to its neighbors, or through the introduction of a local potential.

\emph{Note added} --- The results described here have been originally
presented in reference \onlinecite{Pereira:2006PhD} during 2006. While
preparing this manuscript we became aware of the preprint
\onlinecite{Wu:2007} with
some overlapping results regarding local impurities.


\section{Acknowledgments}

We acknowledge many motivating and fruitfull discussions with 
N.~M.~R.~Peres and F. Guinea.
V.~M.~Pereira is supported by Funda\c{c}\~{a}o para a Ci\^{e}ncia e a Tecnologia
via SFRH/BPD/27182/2006.
V.~M.~Pereira and J.~M.~B.~Lopes~dos~Santos further acknowledge POCI 2010 via
the grant PTDC/FIS/64404/2006. 
A.~H.~Castro~Neto was supported through the NSF grant DMR-0343790.

\bibliographystyle{apsrev}
\bibliography{graphene_disorder}


\begin{acronym}[MOSFET]
  \color{white}
  \acro{BZ}{Brillouin zone}
  \acro{CPA}{coherent potential approximation}
  \acro{DOS}{density of states}
  \acro{HOPG}{highly oriented pyrolytic graphite}
  \acro{IPR}{inverse participation ratio}
  \acro{LDOS}{local density of states}
  \acro{MOSFET}{metal-oxide-semiconductor field effect transistor}
  \acro{QED}{quantum electrodynamics}
  \acro{QHE}{quantum Hall effect}
  \acro{STM}{scanning tunneling microscopy}
  \acro{WS}{Wigner-Seitz}
\end{acronym}

\end{document}